# MIMO Precoding with X- and Y-Codes

Saif Khan Mohammed, Emanuele Viterbo, Yi Hong, and Ananthanarayanan Chockalingam

### Abstract

We consider a time division duplex (TDD) $n_t \times n_r$ multiple-input multiple-output (MIMO) system with channel state information (CSI) at both the transmitter and receiver. We propose X- and Y-Codes to achieve high multiplexing and diversity gains at low complexity. The proposed precoding schemes are based upon the singular value decomposition (SVD) of the channel matrix which transforms the MIMO channel into parallel subchannels. Then X- and Y-Codes are used to improve the diversity gain by pairing the subchannels, prior to SVD precoding. In particular, the subchannels with good diversity are paired with those having low diversity gains. Hence, a pair of channels is jointly encoded using a $2 \times 2$ real matrix, which is fixed *a priori* and does not change with each channel realization. For X-Codes these matrices are 2-dimensional rotation matrices parameterized by a single angle, while for Y-Codes, these matrices are 2-dimensional upper left triangular matrices. The complexity of the maximum likelihood decoding (MLD) for both X- and Y-Codes is low. Specifically, the decoding complexity of Y-Codes is the same as that of a scalar channel. Moreover, we propose X-, Y-Precoders with the same structure as X-, Y-Codes, but the encoding matrices adapt to each channel realization. The optimal encoding matrices for X-, Y-Codes/Precoders are derived analytically. Finally, it is observed that X-Codes/Precoders perform better for well-conditioned channels, while Y-Codes/Precoders perform better for ill-conditioned channels, when compared to other precoding schemes in the literature.

## I. Introduction

We consider time division duplexed (TDD) $n_t \times n_r$ multiple-input multiple-output (MIMO) systems, where channel state information (CSI) is fully available both at the transmitter and receiver. Channels in such systems are typically slowly time varying and therefore subject to block fading. It is known that precoding techniques can provide large performance improvements in such scenarios. A popular precoding approach is based on singular value decomposition (SVD) [1], [2] of the channel so that the MIMO channel can be seen as parallel channels. Waterfilling schemes [3] can be used to further improve the total capacity.

S.K. Mohammed and A. Chockalingam are with Indian Institute of Science, Bangalore 560012, India, E-mails: saifind2007@yahoo.com and achockal@ece.iisc.ernet.in. Y. Hong and E. Viterbo are with DEIS, University of Calabria, Rende CS 87036, Italy, E-mails: yi.winnie.hong@gmail.com and viterbo@deis.unical.it. S.K. Mohammed is currently visiting DEIS, University of Calabria, Italy.

Another approach is based on channel inversion (CI) [4], named as zero-forcing (ZF) precoding, which however suffers from a loss of power efficiency. Non-linear precoding such as Tomlinson-Harashima (TH) precoding [5], [6] was exploited in [7]. Linear precoders, which involve simple linear pre- and post-processing, have been proposed in [8], [9] and references therein. Despite having low encoding and decoding complexity, the linear precoding schemes and the TH precoder have a low diversity order if all the modes of transmission are utilized (i.e., full-rate). Diversity order can be improved, only by transmitting over a subset of all possible modes of transmission, but this results in a rate loss. Precoders based on lattice reduction techniques [10] and vector perturbation [11] can achieve full rate and high diversity, but at the cost of high complexity. We therefore see a tradeoff between rate, diversity and computational complexity. This motivates us to design precoding schemes which achieve both full rate and high diversity at low complexities.

In this paper, we consider SVD precoding for MIMO system, which transforms the MIMO channels into parallel subchannels. At the receiver, maximum likelihood decoding (MLD) can be employed separately for each subchannel. To improve the low diversity order of the SVD precoded system, we propose some simple linear codes prior to SVD precoding. These codes are named X- and Y-Codes due to the structure of the encoder matrix, which enables to flexibly pair subchannels with different diversity orders. Specifically, the subchannels with low diversity orders can be paired together with those having high diversity orders, so that the overall diversity order is improved. The main contributions in this paper are:

1) *X-Codes*: A set of 2-dimensional (2-D) real orthogonal matrices is used to jointly code over pairs of subchannels, without increasing the transmit power. Since the matrices are effectively parameterized with a single angle, the design of X-Codes primarily involves choosing the optimal angle for each pair of subchannels. The angles are chosen *a priori* and do not change with each channel realization. This is why we use the term "Code" instead of "Precoder". Further optimization of angles are based upon minimizing the average error performance. At the receiver, we show that the MLD can be easily accomplished using $n_r$ low complexity 2-D real sphere decoders (SDs) [15]. It is shown that X-Codes have better error performance than that of other precoders, yet it becomes worse when the pair of subchannels is *poorly conditioned* (for definition, see Theorem 2). This motivates us to propose the Y-Codes.

2) *Y-Codes*: Instead of using rotation for pairing subchannels, we use a linear code generator matrix which is upper left triangular. Y-Codes are parameterized with 2 parameters corresponding to power allocated to the two subchannels. These parameters are computed so as to minimize the average error probability. The MLD complexity is the same as that of the scalar channels in linear precoders [8], [9] and is less than that of the X-Codes, while the performance of Y-Codes is better than that of X-Codes for ill-conditioned channel pairs.

3) *X-, Y-Precoders*: The X- and Y-Precoders employ the same pairing structure as that in X-, Y-Codes. However, the code generator matrix for each pair of subchannels is chosen for each channel realization. We observed that the error performance of X- and Y-Precoders is better than that of X- and Y-Codes.

A precoding scheme, named *E-dmin*, has been recently proposed in [12]. The structure of its precoding matrix is similar to that of X-Codes. However, the E-dmin precoder is only optimized for 4-QAM symbols, but suffers from loss in power efficiency with higher order modulation. The performance and decoding complexity of E-dmin are worse when compared to the proposed X-, Y-Codes.

The rest of the paper is organized as follows. Section II introduces the system model and SVD precoding. In Section III we present the pairing of subchannels as a general coding strategy to achieve higher diversity order in fading channels. In Section IV, we propose the X-Codes and the X-Precoders. We show that ML decoding can be achieved with $n_r$ 2-D real SDs. We also analyze the error performance and present the design of optimal X-Codes and X-Precoders. In Section V, we propose the Y-Codes and Y-Precoder. We show that they have very low decoding complexity. We analyze the error performance and derive expressions for the optimal Y-Codes and Y-Precoders. Section VI shows the simulation results and comparisons with other precoders. Section VII discusses the complexity of the X-, Y-Codes/Precoders in comparison with other precoders. Conclusions are drawn in Section VIII.

*Notations*: Superscripts $T$, $\dagger$, and $*$ denote transposition, Hermitian transposition, and complex conjugation, respectively. The $n \times n$ identity matrix is denoted by $\mathbf{I}_n$, and the zero matrix is denoted by $\mathbf{0}$. The $\mathbb{E}[\cdot]$ is the expectation operator, $\|\cdot\|$ denotes the Euclidean norm, and $|\cdot|$ denotes the absolute value of a complex number. The fields of complex numbers, real numbers and non-negative real numbers are $\mathbb{C}$, $\mathbb{R}$ and $\mathbb{R}^+$, respectively. Furthermore, $\lfloor c \rfloor$ denotes the largest integer less than $c$. Finally, we let $\Re(\cdot)$ and $\Im(\cdot)$ denote the real and imaginary parts of a complex argument.

## II. System model and SVD precoding

We consider a TDD $n_t \times n_r$ MIMO ($n_r \leq n_t$), where the channel state information (CSI) is known perfectly at both the transmitter and receiver. Let $\mathbf{x} = (x_1, \ldots, x_{n_t})^T$ be the vector of symbols transmitted by the $n_t$ transmit antennas, and let $\mathbf{H} = \{h_{ij}\}$, $i = 1, \ldots, n_r$, $j = 1, \ldots, n_t$, be the $n_r \times n_t$ channel coefficient matrix, with $h_{ij}$ as the complex channel gain between the $j$-th transmit antenna and the $i$-th receive antenna. The standard Rayleigh flat fading model is assumed with $h_{ij} \sim \mathcal{N}_c(0, 1)$, i.e., i.i.d. complex Gaussian random variables with zero mean and unit variance. The received vector with $n_r$ symbols is given by

$$\mathbf{y} = \mathbf{Hx} + \mathbf{n} \tag{1}$$

where $\mathbf{n}$ is a spatially uncorrelated Gaussian noise vector such that $\mathbb{E}[\mathbf{nn}^\dagger] = N_0 \mathbf{I}_{n_r}$. Such a system has a maximum multiplexing gain of $n_r$.

Let the number of transmitted information symbols be $n_s$ ($n_s \leq n_r$). The information bits are first mapped to the information symbol vector $\mathbf{u} = (u_1, \ldots, u_{n_s})^T \in \mathbb{C}^{n_s}$, which is then mapped to the data symbol vector $\mathbf{z} = (z_1, \ldots, z_{n_s})^T \in \mathbb{C}^{n_s}$ using a $n_s \times n_s$ matrix $\mathbf{G}$.

$$\mathbf{z} = \mathbf{Gu} + \mathbf{u}^0 \tag{2}$$

where $\mathbf{u}^0 \in \mathbb{C}^{n_s}$ is a displacement vector used to reduce the average transmitted power.

Let $\mathbf{T}$ be the $n_t \times n_s$ precoding matrix which is applied to the data symbol vector to yield the transmitted vector

$$\mathbf{x} = \mathbf{Tz} \tag{3}$$

In general $\mathbf{T}, \mathbf{G}$ and $\mathbf{u}_0$ are derived from the perfect knowledge of $\mathbf{H}$ at the transmitter and they are crucial to the system performance and complexity. The transmission power constraint is given by

$$\mathbb{E}[\|\mathbf{x}\|^2] = P_T \tag{4}$$

and we define the signal-to-noise ratio (SNR) as

$$\gamma \triangleq \frac{P_T}{N_0}$$

The proposed X-, Y-Codes are based on the SVD precoding technique, which is based on the singular value decomposition of the channel matrix $\mathbf{H} = \mathbf{U}\mathbf{\Lambda}\mathbf{V}$, where $\mathbf{U} \in \mathbb{C}^{n_r \times n_r}$, $\mathbf{\Lambda} \in \mathbb{C}^{n_r \times n_r}$, $\mathbf{V} \in \mathbb{C}^{n_r \times n_t}$, and $\mathbf{U}\mathbf{U}^\dagger = \mathbf{V}\mathbf{V}^\dagger = \mathbf{I}_{n_r}$, and $\mathbf{\Lambda} = \mathrm{diag}(\lambda_1, \ldots, \lambda_{n_r})$, with $\lambda_1 \geq \lambda_2 \cdots \geq \lambda_{n_r} \geq 0$.

Let $\tilde{\mathbf{V}} \in \mathbb{C}^{n_s \times n_t}$ be the submatrix with the first $n_s$ rows of $\mathbf{V}$. The SVD precoder uses

$$\mathbf{T} = \tilde{\mathbf{V}}^\dagger,$$
$$\mathbf{G} = \mathbf{I}_{n_s},$$
$$\mathbf{u}^0 = \mathbf{0} \tag{5}$$

and the receiver gets

$$\mathbf{y} = \mathbf{H}\mathbf{T}\mathbf{u} + \mathbf{n} \tag{6}$$

Let $\tilde{\mathbf{U}} \in \mathbb{C}^{n_r \times n_s}$ be the submatrix with the first $n_s$ columns of $\mathbf{U}$. The receiver computes

$$\mathbf{r} = \tilde{\mathbf{U}}^\dagger \mathbf{y} = \tilde{\mathbf{\Lambda}}\mathbf{u} + \mathbf{w} \tag{7}$$

where $\mathbf{w} \in \mathbb{C}^{n_s}$ is still an uncorrelated Gaussian noise vector with $\mathbb{E}[\mathbf{w}\mathbf{w}^\dagger] = N_0 \mathbf{I}_{n_s}$, $\tilde{\mathbf{\Lambda}} \triangleq \mathrm{diag}(\lambda_1, \lambda_2, \cdots \lambda_{n_s})$, and $\mathbf{r} = (r_1, \ldots, r_{n_s})^T$. The SVD precoding therefore transforms the channel into $n_s$ parallel channels

$$r_i = \lambda_i u_i + w_i \qquad i = 1, \ldots, n_s \tag{8}$$

with non-negative fading coefficients $\lambda_i$. The overall error performance is dominated by the minimum singular value $\lambda_{n_s}$. When $n_s = n_r = n_t$, the resulting diversity order is only 1.

## III. PAIRING GOOD AND BAD SUBCHANNELS

Without loss of generality, we consider only the full-rate SVD precoding scheme with even $n_r$ and $n_s = n_r$. The matrix $\mathbf{G} \in \mathbb{C}^{n_r \times n_r}$ is now used to pair (jointly encode) different subchannels in order to improve the diversity order of the system. The precoding matrix $\mathbf{T} \in \mathbb{C}^{n_t \times n_r}$ and the transmitted vector $\mathbf{x}$ are given by

$$\mathbf{T} = \mathbf{V}^\dagger, \qquad \mathbf{x} = \mathbf{V}^\dagger(\mathbf{G}\mathbf{u} + \mathbf{u}^0) \tag{9}$$

Let the list of pairings be $(i_k, j_k) \in [1, n_r]$, $k \in [1, n_r/2]$ and $i_k < j_k$. On the $k$-th pair, consisting of subchannels $i_k$ and $j_k$, the information symbols $u_{i_k}$ and $u_{j_k}$ are jointly coded using a $2 \times 2$ matrix $\mathbf{A}_k$. In order to reduce the ML decoding complexity, we restrict the entries of $\mathbf{A}_k$ to be real valued. Each $\mathbf{A}_k \triangleq \{a_{k,i,j}\}$, $i, j \in [1, 2]$, is a submatrix of the code matrix $\mathbf{G}$ as shown below.

$$
\begin{aligned}
g_{i_k, i_k} = a_{k,1,1} \quad & g_{i_k, j_k} = a_{k,1,2} \\
g_{j_k, i_k} = a_{k,2,1} \quad & g_{j_k, j_k} = a_{k,2,2}
\end{aligned}
\tag{10}
$$

where $g_{i,j}$ is the entry of $\mathbf{G}$ in the $i$-th row and $j$-th column.

We shall see later, that an optimal pairing in terms of achieving the best diversity order is one in which the $k$-th subchannel is paired with the $(n_r - k + 1)$-th subchannel. For example, with $n_r = 6$, the X-Code structure is given by

$$
\mathbf{G} =
\begin{bmatrix}
a_{1,1,1} & & & & & a_{1,1,2} \\
& a_{2,1,1} & & & a_{2,1,2} & \\
& & a_{3,1,1} & a_{3,1,2} & & \\
& & a_{3,2,1} & a_{3,2,2} & & \\
& a_{2,2,1} & & & a_{2,2,2} & \\
a_{1,2,1} & & & & & a_{1,2,2}
\end{bmatrix}
\tag{11}
$$

and the Y-Code structure is given by [1]

$$
\mathbf{G} =
\begin{bmatrix}
a_{1,1,1} & & & & & a_{1,1,2} \\
& a_{2,1,1} & & & a_{2,1,2} & \\
& & a_{3,1,1} & a_{3,1,2} & & \\
& & a_{3,2,1} & & & \\
& a_{2,2,1} & & & & \\
a_{1,2,1} & & & & &
\end{bmatrix}
\tag{12}
$$

Let

$$
\mathbf{u}_k \triangleq [u_{i_k}, u_{j_k}]^T
$$

Due to the transmit power constraint in (4), and uniform power allocation between the $n_r/2$ pairs, the encoder matrices $\mathbf{A}_k$ must satisfy

$$
\mathbb{E}\left[\|\mathbf{A}_k \mathbf{u}_k + \mathbf{u}_k^0\|^2\right] = \frac{2P_T}{n_r}
\tag{13}
$$

The expectation in (13) is over the distribution of the information symbol vector $\mathbf{u}_k$. $\mathbf{u}_k^0$ is the subvector of the displacement vector $\mathbf{u}^0$ for the $k$-th pair.

The matrices $\mathbf{A}_k$ for X- and Y- codes can be either fixed *a priori* or can change with every channel realization. The latter case leads to the X-, Y-Precoders.

---

[1] The names X- and Y-Codes are due to the structure of the code generating matrices in (11) and (12).

*A. ML Decoding*

Given the received vector $\mathbf{y}$, the receiver computes

$$\mathbf{r} = \mathbf{U}^\dagger \mathbf{y} - \mathbf{\Lambda} \mathbf{u}^0 \tag{14}$$

Using (1) and (9), we can rewrite (14) as

$$\mathbf{r} = \mathbf{\Lambda}\mathbf{G}\mathbf{u} + \mathbf{w} = \mathbf{M}\mathbf{u} + \mathbf{w} \tag{15}$$

where $\mathbf{M} \triangleq \mathbf{\Lambda}\mathbf{G}$ is the equivalent channel gain matrix and $\mathbf{w} \triangleq \mathbf{U}^\dagger \mathbf{n}$ is a noise vector with the same statistics as $\mathbf{n}$. Further, we let

$$\begin{aligned}
\mathbf{r}_k &\triangleq [r_{i_k}, r_{j_k}]^T \\
\mathbf{w}_k &\triangleq [w_{i_k}, w_{j_k}]^T
\end{aligned}$$

Let $\mathbf{M}_k \in \mathbb{R}^{2\times 2}$ denote the $2 \times 2$ submatrix of $\mathbf{M}$ consisting of entries in the $i_k$ and $j_k$ rows and columns. Then (15) can be equivalently written as

$$\mathbf{r}_k = \mathbf{M}_k \mathbf{u}_k + \mathbf{w}_k, \quad k = 1, \ldots, \frac{n_r}{2} \tag{16}$$

$\Re(\mathbf{u}_k) \in \mathcal{S}_k$, where $\mathcal{S}_k$ is a finite signal set in the 2-dimensional real space .

Assuming that the same set is used for the imaginary component, the spectral efficiency $\eta$ is given by

$$\eta = 2 \sum_{k=1}^{\frac{n_r}{2}} \log_2(|\mathcal{S}_k|) \tag{17}$$

From (16), it is clear that MLD reduces to separate MLDs of the $k$ pairs, which can be further separated into independent ML decoding of the real and imaginary components of $\mathbf{u}_k$. Then the MLD for the $k$-th pair is given by

$$\Re(\hat{\mathbf{u}}_k) = \arg \min_{\Re(\mathbf{u}_k) \in \mathcal{S}_k} \|\Re(\mathbf{r}_k) - \mathbf{M}_k \Re(\mathbf{u}_k)\|^2 \tag{18}$$

and

$$\Im(\hat{\mathbf{u}}_k) = \arg \min_{\Im(\mathbf{u}_k) \in \mathcal{S}_k} \|\Im(\mathbf{r}_k) - \mathbf{M}_k \Im(\mathbf{u}_k)\|^2 \tag{19}$$

where $\hat{\mathbf{u}}_k$ is the output of the ML detector for the $k$-th pair.

*B. Performance Analysis*

Let $P_k$ denote the word error probability (WEP) for the $k$-th pair of sub-channels with the ML receiver. The overall WEP for the transmitted information symbol vector is given by

$$P = 1 - \prod_{k=1}^{n_r/2} (1 - P_k). \tag{20}$$

From (18) and (19), we see that WEPs for the real and the imaginary components of the $k$-th pair are the same. Therefore, without loss of generality we can compute the WEP only for the real component, denoted by $P_k^{'}$, and then $P_k = 1 - (1 - P_k^{'})^2$. Let us further denote by $P_k^{'}(\Re(\mathbf{u}_k))$ the probability of the real part of the ML decoder decoding not in favor of $\Re(\mathbf{u}_k)$ when $\mathbf{u}_k$ is transmitted on the $k$-th pair. $P_k^{'}$ can then be expressed in terms of $P_k^{'}(\Re(\mathbf{u}_k))$ as follows

$$P_k^{'} = \frac{1}{|\mathcal{S}_k|} \sum_{\Re(\mathbf{u}_k)} P_k^{'}(\Re(\mathbf{u}_k)) \tag{21}$$

where $P_k^{'}(\Re(\mathbf{u}_k))$ has to be evaluated differently for X-, Y-Codes and X-, Y-Precoders. To explain this difference we need the following definitions.

For a given channel realization, i.e., deterministic value of $\lambda_{i_k}$ and $\lambda_{j_k}$ for the $k$-th pair, we let $P_k^{'}(\Re(\mathbf{u}_k), \lambda_{i_k}, \lambda_{j_k}, \mathbf{A}_k)$ be the error probability of MLD for the real component of the $k$-th channel, given that the information symbol $\mathbf{u}_k$ was transmitted on the $k$-th pair. For X-, Y-Codes, the matrices $\mathbf{A}_k$ are fixed *a priori* and are not function of the deterministic value of channel gains, and therefore, $P_k^{'}(\Re(\mathbf{u}_k))$, is given by

$$P_k^{'}(\Re(\mathbf{u}_k)) = \mathbb{E}_{(\lambda_{i_k}, \lambda_{j_k})} \left[ P_k^{'}(\Re(\mathbf{u}_k), \lambda_{i_k}, \lambda_{j_k}, \mathbf{A}_k) \right] \tag{22}$$

We observe that $P_k^{'}(\Re(\mathbf{u}_k))$ is actually a function of $\mathbf{A}_k$ and therefore the optimal error performance is obtained by minimizing (21) over $\mathbf{A}_k$. Then the optimal matrix for the $k$-th pair is given by

$$\mathbf{A}_k^{opt} = \arg \min_{\mathbf{A}_k} \sum_{\Re(\mathbf{u}_k)} \mathbb{E}_{(\lambda_{i_k}, \lambda_{j_k})} \left[ P_k^{'}(\Re(\mathbf{u}_k), \lambda_{i_k}, \lambda_{j_k}, \mathbf{A}_k) \right] \tag{23}$$

The minimization in (23) is constrained over matrices $\mathbf{A}_k$ which satisfy (13). The optimal error performance $P_k^{opt}$ is given by

$$P_k^{opt} = \frac{1}{|\mathcal{S}_k|} \sum_{\Re(\mathbf{u}_k)} \mathbb{E}_{(\lambda_{i_k}, \lambda_{j_k})} \left[ P_k^{'}(\Re(\mathbf{u}_k), \lambda_{i_k}, \lambda_{j_k}, \mathbf{A}_k^{opt}) \right] \tag{24}$$

For the X-, Y-Precoder, the matrices $\mathbf{A}_k$ are chosen every time the channel changes. For optimal performance, the matrices $\mathbf{A}_k$ are chosen so as to minimize the error probability for a given channel realization Let $\mathbf{A}_k^{opt}$, the optimal encoding matrix for the $k$th pair is then given by

$$\mathbf{A}_k^{opt}(\lambda_{i_k}, \lambda_{j_k}) = \arg \min_{\mathbf{A}_k} \sum_{\Re(\mathbf{u}_k)} P_k^{'}(\Re(\mathbf{u}_k), \lambda_{i_k}, \lambda_{j_k}, \mathbf{A}_k) \tag{25}$$

The optimal error performance for X-, Y-Precoders is therefore given by

$$P_k^{opt} = \frac{1}{|\mathcal{S}_k|} \sum_{\Re(\mathbf{u}_k)} \mathbb{E}_{(\lambda_{i_k}, \lambda_{j_k})} \left[ P_k^{'}(\Re(\mathbf{u}_k), \lambda_{i_k}, \lambda_{j_k}, \mathbf{A}_k^{opt}(\lambda_{i_k}, \lambda_{j_k})) \right] \tag{26}$$

Comparing (26) and (24), we immediately observe that the optimal error performance of X-, Y-Precoders is better than that of X-, Y-Codes. Our next goal is to derive an analytic expression for $P_k^{'}(\Re(\mathbf{u}_k))$. We shall only discuss the derivation for X-, Y-Codes, since the performance of X-, Y-Precoders is better than X-, Y-Codes and therefore have at least as much diversity order as X-, Y-Codes. Getting an exact analytic expression is difficult, and therefore we try to get tight upper bounds using the union bound.

**Theorem 1:** The upper bound to $P_k^{'}$ is given by

$$P_k^{'} \leq c_k(|\mathcal{S}_k| - 1) \left( \frac{\gamma g_k(\mathbf{A}_k)}{2P_T} \right)^{-\delta_k} + o(\gamma^{-\delta_k}) \tag{27}$$

where

$$\delta_k \quad \triangleq \quad (n_t - i_k + 1)(n_r - i_k + 1)$$

$$c_k \quad \triangleq \quad \frac{C(i_k)((2\,\delta_k - 1) \quad \cdots \quad 5 \cdot 3 \cdot 1)}{2\delta_k}$$

where $g_k(\mathbf{A}_k)$ is the *generalized minimum distance*, as defined in (63) (See Appendix A), $C(m)$ ($1 \le m \le \min(n_r, n_t)$) is defined in [17].

*Proof* – See Appendix A. ∎

Let us define the *overall diversity order*

$$\delta_{ord} \triangleq \lim_{\gamma \to \infty} \frac{-\log P}{\log \gamma}$$

Then, it is obvious that

$$\delta_{ord} \ge \min_k \delta_k. \tag{28}$$

This bound also holds for the X-, Y-Precoders, since the error performance of the X-, Y-Precoders is always better than that of X-, Y-Codes.

## C. Design of Optimal Pairing

From the lower bound on $\delta_{ord}$ (given by (28)) it is clear that the following pairing of sub-channels

$$i_k = k\,, \ j_k = n_r - k + 1. \tag{29}$$

achieves the following best lower bound

$$\delta_{ord} \ge \left(\frac{n_r}{2} + 1\right)\left(n_t - \frac{n_r}{2} + 1\right) \tag{30}$$

*Remark 1:* Note that this corresponds to a cross-form generator matrix $\mathbf{G}$, and is not the only pairing for the best lower bound. Also we note that the diversity order improves significantly, when compared to the case of no pairing. It can be shown that, if only $n_s$ ($n_s$ is even) out of the $n_r$ subchannels are used for transmission, the lower bound on the achievable diversity order is $(n_r - \frac{n_s}{2} + 1)(n_t - \frac{n_s}{2} + 1)$. ∎

Although it is hard to compute $\mathbf{A}_k^{opt}$, we can compute the best $\mathbf{A}_k$, denoted by $\mathbf{A}_k^*$, which minimizes the upper bound on $P_k'$ in (27). Then we have

$$\mathbf{A}_k^* = \arg \max_{\mathbf{A}_k \,|\, \mathbb{E}\left[\|\mathbf{A}_k \mathbf{u}_k + \mathbf{u}_k^0\|^2\right] = \frac{2P_T}{n_r}} \frac{g_k(\mathbf{A}_k)}{2P_T} \tag{31}$$

Using (27), (29) and (31), we obtain

$$P_k' \le c_k(|\mathcal{S}_k| - 1)\left(\frac{\gamma\, g_k(\mathbf{A}_k^*)}{2P_T}\right)^{-\delta_k^*} + o(\gamma^{-\delta_k^*}) \tag{32}$$

where $\delta_k^* \triangleq (n_t - k + 1)(n_r - k + 1)$.

## IV. X-Codes and X-Precoders

### A. X-Codes and X-Precoders: Encoding

For X-Codes, each symbol in $\mathbf{u}$ takes values from a regular $M^2$-QAM constellation, which consists of the $M$-PAM constellation $\mathcal{S} \triangleq \{\tau(2i - (M-1)) \mid i = 0, 1, \cdots (M-1)\}$ used in quadrature on the real and the imaginary components of the channel. The constant

$$\tau \triangleq \sqrt{\frac{3E_s}{2(M^2-1)}}$$

and

$$E_s = \frac{P_T}{n_r}$$

is the average symbol energy for each information symbol in the vector $\mathbf{u}$. Gray mapping is used to map the bits separately to the real and imaginary component of the symbols in $\mathbf{u}$. We fix $\mathbf{u}^0$ to be the zero vector. In order to avoid transmitter power enhancement, we impose an orthogonality constraint on each $\mathbf{A}_k$ and parameterize it with a single angle $\theta_k$.

$$\mathbf{A}_k = \begin{bmatrix} \cos(\theta_k) & \sin(\theta_k) \\ -\sin(\theta_k) & \cos(\theta_k) \end{bmatrix} \quad k = 1, \ldots n_r/2 \tag{33}$$

We notice that 1) both $\mathbf{A}_k$ and $\mathbf{G}$ are orthogonal; 2) for X-Codes we fix the angles $\theta_k$ *a priori* whereas for the X-Precoders we change the angles for each channel realization.

### B. X-Codes and X-Precoders: ML Decoding

From (18) and (19) it is obvious that two 2-D real SDs are needed for each pair. Since there are $\frac{n_r}{2}$ pairs, the total decoding complexity is $n_r$ 2-D real SDs. For X-Codes, the matrices $\mathbf{M}_k$ in (18) and (19) are given by

$$\mathbf{M}_k = \begin{bmatrix} \lambda_{i_k} \cos(\theta_k) & \lambda_{i_k} \sin(\theta_k) \\ -\lambda_{j_k} \sin(\theta_k) & \lambda_{j_k} \cos(\theta_k) \end{bmatrix} \tag{34}$$

### C. Optimal design of X-Codes

In order to find the best angle $\theta_k$ for the $k$-th pair, we attempt to maximize $g_k(\mathbf{A}_k)$ under the transmit power constraints. For X-Codes, let $\mathbf{z}_k \triangleq \Re(\mathbf{u}_k) - \Re(\mathbf{v}_k)$ be the difference vector between any two information vectors, which can be written as

$$\mathbf{z}_k = \sqrt{\frac{6E_s}{(M^2-1)}}(p, q)^T \qquad (p, q) \in \mathbb{S}_M$$

where

$$\mathbb{S}_M \triangleq \{(p, q) | 0 \leq |p| \leq (M-1), 0 \leq |q| \leq (M-1), (p, q) \neq (0, 0)\}$$

As defined in Appendix A, the *generalized pairwise distance* is

$$\tilde{d}_k^2(\Re(\mathbf{u}_k), \Re(\mathbf{v}_k), \mathbf{A}_k) = \frac{6P_T \left(p \cos(\theta_k) + q \cos(\theta_k)\right)^2}{n_r(M^2-1)} \tag{35}$$

Since $\mathbf{A}_k$ is parameterizable with a single angle $\theta_k$, let

$$g_k(\theta_k, M) \triangleq g_k(\mathbf{A}_k)$$

Also let

$$\varphi_{p,q} \triangleq \tan^{-1}\left(\frac{q}{p}\right)$$

Using (35) and the (63), we have

$$g_k(\theta_k, M) = \frac{6P_T \min_{(p,q)\in\mathbb{S}_M}(p^2+q^2)\cos^2(\theta_k-\varphi_{p,q})}{n_r(M^2-1)}$$

Using (31), the best $\theta_k$, denoted by $\theta_k^*$, is given by

$$\theta_k^* = \arg\max_{\theta_k\in[0,\frac{\pi}{4}]}\min_{(p,q)\in\mathbb{S}_M}(p^2+q^2)\cos^2(\theta_k-\varphi_{p,q}) \tag{36}$$

Following (32), the best achievable upper bound for $P_k^{'}$ is

$$P_k^{'} \le (M^2-1)c_k\left(\frac{3\gamma g_k(\theta_k^*, M)}{n_r(M^2-1)}\right)^{-\delta_k^*} + o(\gamma^{-\delta_k^*}) \tag{37}$$

*Remark 2:* It is easily shown by the symmetry of the set $\mathbb{S}_M$ that it suffices to consider $\theta_k \in [0, \frac{\pi}{4}]$ for the maximization in (36). The min-max optimization problem does not have explicit analytical solutions except for small values of $M$, for example $M = 2$. But since the encoder matrices are fixed *a priori*, these computations can be performed off-line. ∎

For small MIMO systems (for example $2 \times 2$ MIMO) it is possible to get a tighter upper bound by evaluating the expectation in (61) (See Appendix A). $P_1^{'}$ is then upper bounded as follows.

$$P_1^{'} \le \sum_{(p,q)\in\mathbb{S}_M} \frac{(70/81)(M^2-1)^4\gamma^{-4}}{M^2(p\cos(\theta_1)+q\sin(\theta_1))^6(p^2+q^2)} + o(\gamma^{-4}) \tag{38}$$

where $\theta_1$ is the angle used for the only pair. For larger MIMO systems it is preferable to use the inequality in (62) (See Appendix A), since evaluating the expectation containing two singular values is tedious. In Fig. 1, we compare the word error probability of a $2 \times 2$ MIMO system with that given by (38), and observe that the union bound is indeed tight at high SNR.

In Fig. 2, we plot the variation of the upper bound to the WEP w.r.t. the angle $\theta_1$ for the $2 \times 2$ MIMO system with 4-QAM and 16-QAM modulation. We observe that WEP is indeed sensitive to the rotation angle. With 4-QAM modulation, the WEP worsens as the angle approaches either 0 or 45 degrees. With 16-QAM modulation, the performance is even more sensitive to the rotation angle. Moreover we observe that the performance is poor, when the angles are chosen near 18.5, 26.6 and 33.7 degrees, corresponding to $\varphi_{3,1}$, $\varphi_{2,1}$, and $\varphi_{3,2}$, respectively. From (61), it is clear that the performance at high SNR is determined by the minimum value of the distance $\|\mathbf{M}_k(\Re(\mathbf{u}_k)-\Re(\mathbf{v}_k))\|^2$, which is

$$(p^2+q^2)\left(\lambda_{i_k}^2\cos^2(\theta_k-\varphi_{p,q})+\lambda_{j_k}^2\sin^2(\theta_k-\varphi_{p,q})\right)$$

when $(p, q)$ takes values over the set $\mathbb{S}_M$. If $\theta_k = \tan^{-1}(-p/q)$ for some $(p, q) \in \mathbb{S}_M$, then the minimum distance is independent of $\lambda_{i_k}$ and depends only upon $\lambda_{j_k}$. This implies a loss of diversity order since the diversity order of the square fading coefficient $\lambda_{j_k}^2$ is less than that of $\lambda_{i_k}^2$. For the case of $n_r = n_t = 2$, this would mean a reduction of diversity order from $4$ to $1$. The set $\mathbb{S}_M$ and the critical angles are illustrated in Fig. 3.

### D. Optimal design of X-Precoder

For X-Precoders, the optimal rotation angle is tedious to compute due to lack of exact expressions for error probability. Just like X-Codes we resort to bounds on error performance. It is possible to get union bound expression for the error probability of the $k$-th pair. However, we do not further upper bound the union bound by using (62), since by doing so we would have lost information about $\lambda_{j_k}$. Instead in the pairwise sum, we look for the term with the highest contribution to the union bound and try to minimize this term. The best angle for the $k$-th pair is then given by,

$$
\begin{aligned}
\tilde{\theta}_k(\lambda_{i_k}, \lambda_{j_k}) &= \arg \max_{\theta_k \in [0, 2\pi]} \min_{(p,q) \in \mathbb{S}_M} d_k^2(p, q, \theta_k) \\
&= \arg \max_{\theta_k \in [0, \frac{\pi}{4}]} \min_{(p,q) \in \mathbb{S}_M} d_k^2(p, q, \theta_k)
\end{aligned}
\tag{39}
$$

where

$$
d_k^2(p, q, \theta_k) \triangleq (p^2 + q^2)(\lambda_{i_k}^2 \cos^2(\theta_k - \varphi_{p,q}) + \lambda_{j_k}^2 \sin^2(\theta_k - \varphi_{p,q}))
\tag{40}
$$

Just like for X-Codes, it can be shown that for the maximization in (39), it suffices to consider the range $[0, \frac{\pi}{4}]$ for $\theta_k$. The optimization problem in (39) is difficult, but can be solved exactly for small values of $M$. Also, the minimization over $(p, q) \in \mathbb{S}_M$ need not be over the full set containing $|\mathbb{S}_M| = 4M(M-1)$ elements. In fact it can be shown that the number of elements to searched is at most $(M^2 - 3M + 6)/2$. Therefore, for $M = 4$ (16-QAM), we need not search over the full set of $48$ elements, but rather it suffices to search over only $5$ elements.

**Theorem 2:** For $M = 2$ (4-QAM), the exact $\tilde{\theta}_k(\lambda_{i_k}, \lambda_{j_k})$ is given by,

$$
\begin{cases}
\pi/4 & \beta_k \leq \sqrt{3} \\
\tan^{-1}\left[(\beta_k^2 - 1) - \sqrt{((\beta_k^2 - 1)^2 - \beta_k^2)}\right] & \beta_k > \sqrt{3}
\end{cases}
\tag{41}
$$

where

$$
\beta_k \triangleq \frac{\lambda_{i_k}}{\lambda_{j_k}}.
$$

*Proof* – See Appendix B. ∎

Further let

$$
\tilde{d}_{k,\min}^2(\lambda_{i_k}, \lambda_{j_k}) \triangleq \max_{\theta_k \in [0, \frac{\pi}{4}]} \min_{(p,q) \in \mathbb{S}_M} d_k^2(p, q, \theta_k)
\tag{42}
$$

Using (42), the union bound to $P_k'$ is given by

$$
P_k' \leq (M^2 - 1)\mathbb{E}\left[Q\left(\sqrt{\frac{\tilde{d}_{k,\min}^2(\lambda_{i_k}, \lambda_{j_k})}{2N_0}}\right)\right]
\tag{43}
$$

The expectation in (43) is over the joint distribution of $(\lambda_{i_k}, \lambda_{j_k})$ and is difficult to compute analytically. We therefore use Monte-Carlo simulations to evaluate the exact error probability.

## V. Y-Codes and Y-Precoder

### A. Motivation

It is observed that the error performance at high SNR is dependent on the minimum value of the distance $d_k^2(\Re(\mathbf{u}_k), \Re(\mathbf{v}_k), \mathbf{A}_k)$ over all possible information vectors $\mathbf{u}_k \neq \mathbf{v}_k$. Using the definition for $d_k^2(\Re(\mathbf{u}_k), \Re(\mathbf{v}_k), \mathbf{A}_k)$ (see Appendix A) we have

$$
\begin{aligned}
d_k^2(\Re(\mathbf{u}_k), \Re(\mathbf{v}_k), \mathbf{A}_k) &= \|\mathbf{M}_k(\Re(\mathbf{u}_k) - \Re(\mathbf{v}_k))\|^2 \\
&= \lambda_{i_k}^2 e_{k,1}^2 + \lambda_{j_k}^2 e_{k,2}^2
\end{aligned}
\tag{44}
$$

where $\mathbf{e_k} \triangleq \mathbf{A}_k(\Re(\mathbf{u}_k) - \Re(\mathbf{v}_k))$.

Let $\beta_k$ be the condition number of the equivalent channel for the $k$-th pair (see Theorem 2). We have $\beta_k \geq 1$ since $\lambda_{i_k} \geq \lambda_{j_k}$. For the special case of $\beta_k = 1$, $d_k^2(\Re(\mathbf{u}_k), \Re(\mathbf{v}_k), \mathbf{A}_k)$ is proportional to $\|\mathbf{e}_k\|^2$, which is the Euclidean distance between the code vectors. Therefore, the design of good codes is independent of the channel gain. In such a scenario, it is known that for large $M$ choosing the code vectors as points of the 2-dimensional hexagonal lattice would yield codes with good error performance. However, the design of good codes becomes difficult for values of $\beta_k > 1$. We immediately notice, that the effective Euclidean distance in (44) gives more weight to $e_{k,1}^2$, which is the difference of the vectors along the first component (since $\lambda_{i_k} > \lambda_{j_k}$). Since the total transmit power is constrained, codes should be designed such that the minimum separation of any two code vectors is more along the first component.

For X-Codes and X-Precoder, the minimum separation was increased by rotating the QAM constellation by an optimal angle. However, with this approach, apart from gaining separation along the first component, we also achieve separation along the second component. It is noted that the same diversity order can be achieved even if the minimum separation along the second component is small. Since the average transmit power is constrained, optimal code design would try to choose code vectors such that for the same transmit power, more separation is achieved along the first component (without caring much about the separation achieved along the second component). This observation along with the motivation of further reducing the decoding complexity leads to the design of Y-Codes and Y-Precoders.

### B. Y-Codes and Y-Precoders: Encoding

The matrices $\mathbf{A}_k$ have the structure

$$
\mathbf{A}_k = \begin{bmatrix} a_k & 2a_k \\ 2b_k & 0 \end{bmatrix}
\tag{45}
$$

where $a_k, b_k \in \mathbb{R}^+$. Let $\mathcal{S}_k$ be the set of pairs of integers defined by the Cartesian product

$$
\mathcal{S}_k \triangleq \left\{ [0,1] \times \left[ 0, \dots, \frac{M}{2} - 1 \right] \right\}
$$

For example, with $M = 4$, the set $\mathcal{S}_k$ is given by

$$
\mathcal{S}_k = \left\{ [0,0]^T, \ [0,1]^T, \ [1,0]^T, \ [1,1]^T \right\}
\tag{46}
$$

We consider the 2-D codebook of cardinality $M$ generated by applying $\mathbf{A}_k$ to the elements of $\mathcal{S}_k$. The code vectors $Y_k(v)$, $v = 1, \ldots, M$, are given by

$$Y_k(v) = \left[ a_k \left( (v-1) - \frac{M-1}{2} \right), \quad b_k(-1)^v \right]^T \tag{47}$$

The real and imaginary components of the displacement vector for the $k$-th pair, $\mathbf{u}_k^0$ are given by

$$\Re(\mathbf{u}_k^0) = \Im(\mathbf{u}_k^0) = \left[ -\frac{(M-1)a_k}{2}, \, -b_k \right]^T \tag{48}$$

Due to the transmit power constraint in (13), $a_k$ and $b_k$ must satisfy

$$b_k^2 + a_k^2 \frac{M^2-1}{12} = \frac{P_T}{n_r} \tag{49}$$

Information bits are Gray mapped to codebook indices in such a way. Hence the Hamming distance between bit vectors corresponding to close by code vectors is as small as possible. The only difference between Y-Codes and Y-Precoders is that, for Y-Codes, the parameters $a_k$ and $b_k$ are fixed *a priori*, whereas, for the Y-Precoders, these are chosen every time the channel changes.

### C. Y-Codes and Y-Precoders: ML Decoding

Using our codebook notation, the ML decoding rule in (18) and (19), can be equivalently written as

$$\begin{aligned}
\widehat{v}_k^{(I)} &= \arg \min_{v \in \{0, \cdots, (M-1)\}} \| \Re(\mathbf{r}_k) - \mathbf{\Lambda}_k Y_k(v) \|^2 \\
\widehat{v}_k^{(Q)} &= \arg \min_{v \in \{0, \cdots, (M-1)\}} \| \Im(\mathbf{r}_k) - \mathbf{\Lambda}_k Y_k(v) \|^2
\end{aligned} \tag{50}$$

where $\widehat{v}_k^{(I)}$ and $\widehat{v}_k^{(Q)}$ are ML estimates of the codeword indices transmitted on the real and imaginary components for the $k$-th pair.

We next discuss a low complexity algorithm for the optimization problem in (50). The algorithm is the same for all pairs, and the same for both the real and imaginary components of each pair. Therefore, we only discuss the algorithm for the real component. We first partition the 2-D received signal space $(\mathbb{R}^2)$ into $\left(\frac{M}{2}+1\right)$ regions as follows.

$$\begin{aligned}
R_0 &: \left\{ [x,y]^T \in \mathbb{R}^2 \, | -\infty \leq \left( \frac{x}{\lambda_{i_k} a_k} + \frac{M-1}{2} \right) \leq 1 \right\} \\
R_{\frac{M}{2}} &: \left\{ [x,y]^T \in \mathbb{R}^2 \, | (M-1) \leq \left( \frac{x}{\lambda_{i_k} a_k} + \frac{M-1}{2} \right) \leq +\infty \right\} \\
R_i &: \left\{ [x,y]^T \in \mathbb{R}^2 \, | (2i-1) \leq \left( \frac{x}{\lambda_{i_k} a_k} + \frac{M-1}{2} \right) \leq (2i+1) \right\}
\end{aligned} \tag{51}$$

where $i \in [1, M/2 - 1]$.

In Fig. 4, we illustrate the 5 regions with $M = 8$ for the real component of the $k$-th pair. We next discuss a low complexity ML decoding algorithm for Y-Codes. The first step of the decoding algorithm is to find the region to which the received vector belongs. Let

$$t_k = \left\lfloor \frac{\Re(r_{i_k})}{2\lambda_{i_k} a_k} + \frac{M+1}{4} \right\rfloor$$

The received vector belongs to the region $R_{\zeta_k}$, where $\zeta_k$ is explicitly given by

$$
\zeta_k = \begin{cases} 0 & t_k \leq 0 \\ \frac{M}{2} & t_k \geq \frac{M}{2} \\ t_k & \text{otherwise} \end{cases} \tag{52}
$$

For example, in Fig. 4, the received vectors $\mathbf{p}_1$, $\mathbf{p}_2$, and $\mathbf{p}_3$ belong to $R_0$, $R_1$, and $R_3$ respectively. It can be shown that, once the received vector is decoded to the region $R_{\zeta_k}$, the ML code vector is one among a reduced set of at most 3 code vectors. Therefore, at most 3 Euclidean distances need to be computed to solve the ML detection problem in (50), as compared to computing all the $M$ Euclidean distances in case of a brute force search. For example, in Fig. 4, for the received vector $\mathbf{p}_3 \in R_3$, the ML code vector is among $Y_k(6)$, $Y_k(7)$ or $Y_k(8)$.

However, once we know the region of the received vector, it is possible to directly find the ML code vector even without computing the 3 Euclidean distances. This involves just checking a few linear relations between the 2 components of the received vector. Therefore, the ML decoding complexity of Y-Codes is the same as that of a scalar channel. For example, in Fig. 4, the received vector $\mathbf{p}_3$ is to the right of the perpendicular bisector between $Y_k(6)$ and $Y_k(8)$. The vector $\mathbf{p}_3$ is also above the perpendicular bisector between $Y_k(7)$ and $Y_k(8)$. From these two checks, it can be easily concluded that the ML code vector is $Y_k(8)$. Due to the structure of the codebook, the ML decision regions can be very easily outlined. In Fig. 4, the dotted lines demarcate the boundary of the ML decision regions. The hatched region illustrates the ML decision region of $Y_k(5)$.

### D. Optimal design of Y-Codes

Given the optimal pairing in (29), the next step towards designing optimal Y-Codes is to find the optimal value of $(a_k, b_k)$, which minimizes the average error probability. For Y-Codes, once chosen, $(a_k, b_k)$ are fixed and do not change with every channel realization. Since the ML decision regions are known precisely, it is possible to calculate the exact error probability. With our new codebook notation, we identify code vectors by their index in the codebook, and given by

$$
P_k^{'} = \frac{1}{M} \sum_v P_k^{'}(v) \tag{53}
$$

where $P_k^{'}(v)$ is the probability of error when the code vector $Y_k(v)$ is transmitted. $P_k^{'}(v)$ is given by

$$
P_k^{'}(v) = \begin{cases} \mathbb{E}[g_1(a_k, b_k)] & 3 \leq v \leq (M-2) \\ \mathbb{E}[g_2(a_k, b_k)] & v = 1, M \\ \mathbb{E}[g_1(a_k, b_k) - g_3(a_k, b_k)] & v = 2, (M-1) \end{cases}
$$

The expectation is over the joint distribution of $(\lambda_{i_k}, \lambda_{j_k})$. Let

$$
\Psi_k(x) \triangleq \frac{\sqrt{2}(2a_k\lambda_{i_k}x - a_k^2\lambda_{i_k}^2 - 4b_k^2\lambda_{j_k}^2)}{4b_k\lambda_{j_k}\sqrt{N_0}}
$$

$$
\Phi_k(x) \triangleq -\frac{\sqrt{2}(2a_k\lambda_{i_k}x + a_k^2\lambda_{i_k}^2 + 4b_k^2\lambda_{j_k}^2)}{4b_k\lambda_{j_k}\sqrt{N_0}}
$$

The functions $g_1(a_k, b_k)$, $g_2(a_k, b_k)$ and $g_3(a_k, b_k)$ are given by

$$g_1(a_k, b_k) \triangleq 1 - \int_0^{\lambda_{i_k} a_k} \frac{2e^{\frac{-x^2}{N_0}}}{\sqrt{\pi N_0}} Q\left(\Psi_k(x)\right) dx$$

$$g_2(a_k, b_k) \triangleq 1 - \int_{-\infty}^{\lambda_{i_k} a_k} \frac{e^{\frac{-x^2}{N_0}}}{\sqrt{\pi N_0}} Q\left(\Psi_k(x)\right) dx$$

$$g_3(a_k, b_k) \triangleq \int_{-\infty}^{-\lambda_{i_k} a_k} \frac{e^{\frac{-x^2}{N_0}}}{\sqrt{\pi N_0}} Q\left(\Phi_k(x)\right) dx \tag{54}$$

To compute the optimal $(a_k, b_k)$, we have to minimize $P_k^{'}$ w.r.t. $(a_k, b_k)$ subject to the transmit power constraint in (49). However, it is difficult to get closed form expressions for the optimal $(a_k, b_k)$ due to the intractability of the integrals in (54). This difficulty is further compounded due to the evaluation of expectation over the joint distribution of $(\lambda_{i_k}, \lambda_{j_k})$. However, since $(a_k, b_k)$ are fixed *a priori*, it is always possible to approximately compute the optimal $(a_k, b_k)$ off-line, using Monte-Carlo techniques.

### E. Optimal design of Y-Precoder

For the Y-Precoder, finding the optimal $(a_k, b_k)$ for each channel realization is again difficult due to the intractability of the integrals in (54). In the case of Y-Codes these could be computed offline since $(a_k, b_k)$ are fixed *a priori*. However, for Y-Precoders these cannot be computed offline, since the optimal $(a_k, b_k)$ have to be computed every time the channel changes. Therefore, we try to optimize $(a_k, b_k)$ by minimizing the union bound for $P_k^{'}$. The union bound is given by

$$P_k^{'} \leq (M-1)\mathbb{E}\left[Q\left(\sqrt{\frac{d_{k,\min}^2(a_k, b_k)}{2N_0}}\right)\right] \tag{55}$$

where the expectation is over the joint distribution of $(\lambda_{i_k}, \lambda_{j_k})$ and

$$d_{k,\min}^2(a_k, b_k) \triangleq \min_{v \neq w} \left(\lambda_{i_k}^2 a_k^2 (v-w)^2 + \lambda_{j_k}^2 b_k^2 ((-1)^v - (-1)^w)^2\right) \tag{56}$$

where $v$ and $w$ are distinct indices of the codebook.

The optimal choice of $(a_k, b_k)$, denoted by $(a_k^*, b_k^*)$, which maximizes $d_{k,\min}^2(a_k, b_k)$ for the fixed channel gain of $(\lambda_{i_k}, \lambda_{j_k})$, is given by

**Theorem 3:** The optimal value of $(a_k, b_k)$ defined as

$$(a_k^*, b_k^*) \triangleq \arg \max_{\left\{(a_k, b_k) \in (\mathbb{R}^+)^2 \mid b_k^2 + a_k^2 \frac{M^2-1}{12} = \frac{P_T}{n_r}\right\}} d_{k,\min}^2(a_k, b_k) \tag{57}$$

is given by

$$(a_k^*, b_k^*) = \begin{cases} \left(\sqrt{\frac{12P_T}{n_r(M^2-1)}}, 0\right) & \beta_k^2 \geq \frac{M^2-1}{3} \\ \left(\sqrt{\frac{4P_T}{3n_r(\beta_k^2 + M')}}, \beta_k\sqrt{\frac{P_T}{n_r(\beta_k^2 + M')}}\right) & \beta_k^2 < \frac{M^2-1}{3} \end{cases} \tag{58}$$

where $M' = \frac{M^2-1}{9}$.

The corresponding optimal value of $d^2_{k,\min}(a_k, b_k)$ is given by

$$
d^2_{k,\min}(a^*_k, b^*_k) = \begin{cases} \dfrac{12 P_T \lambda^2_{i_k}}{n_r(M^2-1)} & \beta^2_k \geq \dfrac{M^2-1}{3} \\[3ex] \dfrac{16 P_T \lambda^2_{i_k}}{n_r\left(3\beta^2_k + \dfrac{(M^2-1)}{3}\right)} & \beta^2_k < \dfrac{M^2-1}{3} \end{cases} \tag{59}
$$

*Proof* – See Appendix C. ∎

If we now look back at the codebook for Y-Precoders, we notice that there is power allocation on the 2 channels through the parameters $a_k$ and $b_k$, which can be chosen optimally based upon the knowledge of channel gains. From (58), we observe that the Y-Precoders use only the first channel (the better channel), when channel condition is bad $\left(\beta^2_k \geq \frac{M^2-1}{3}\right)$. For good channel condition, power is distributed between the two channels depending on the channel condition. This adaptive nature of the Y-Precoders enables it to achieve better error performance in badly conditioned channels. Y-Codes also have a fixed-rate allocation between the two channels of a pair, since out of the $\log_2(M)$ bits, one bit can be used to decide whether the vector in the codebook is at even index (corresponding to the second component equal to $+b_k$) or at odd index (corresponding to the second component equal to $-b_k$). The remaining bits are then used to appropriately choose among the vectors at even or odd indices. Therefore, in a way, the proposed Y-Codes always transmits 1 bit of information on the bad channel and $\log_2(M)-1$ bits on the good channel. This rate allocation may not be the best and therefore even better code books can be constructed. One more aspect that is important is the decoding complexity which for the proposed scheme is low and is independent of $M$. It would be challenging to obtain good code books with variable rate allocation and low decoding complexity. We however do not address this problem in this paper.

## VI. SIMULATION RESULTS

In this section, we compare the performance of X-, Y-Codes and X-, Y-Precoders with other precoders. For all the simulations we assume $n_r = n_t$. The optimal matrices $\mathbf{A}_k$ are chosen as discussed previously. Comparisons are made with 1) the E-dmin (equal dmin precoder proposed in [12]), 2) the Arithmetic mean BER precoder (ARITH-MBER) proposed in [8], 3) the Equal Energy linear precoder (EE) based upon optimizing the minimum eigenvalue for a given transmit power constraint [9], 4) the TH precoder based upon the idea of Tomlinson-Harashima precoding applied in the MIMO context [7]) and 5) the channel inversion (CI) known as Zero Forcing precoder [4].

### A. Effect of channel condition on error performance

In Fig. 5, we plot the error performance of all precoding schemes for a $2 \times 2$ MIMO system at $\gamma = 26$ dB, as a function of the condition number $\beta = \lambda_1/\lambda_2$. We fix the total channel gain to be 1, i.e., $\lambda_1{}^2 + \lambda_2{}^2 = 1$, and the target spectral efficiency to be $\eta = 8$ bps/Hz. We briefly discuss the precoding schemes which are compared to the proposed X-, Y-Codes. ARITH-MBER transmits $n_s$ symbols, each from a QAM modulation alphabet. When $n_s = 1$, 256-QAM modulation (16-PAM on the real and imaginary component) is used on the first component of the code vector and the second component is not used for transmission. When $n_s = 2$, 16-QAM modulation is used on both the components. E-dmin is a precoding scheme in which the complex linear precoding matrix is adapted to each channel realization, but both the channels are always used (i.e. $n_s = 2$). The modulation alphabet is 16-QAM.

In Fig. 5, we notice that for schemes, which are fixed and do not adapt with the varying channel, have good error performance for small values of $\beta$. The BER performance is however poor with increasing $\beta$. Error performance of X-Codes is also seen to deteriorate with increasing $\beta$. The only exception are the Y-Codes and ARITH-MBER. For ARITH-MBER with $n_s = 1$ the opposite is true since it always uses only 1 channel for transmission. The performance of Y-Codes is more stable with increasing $\beta$ due to the fact that the codebook is designed in such a way to maximize the minimum separation along the first component without caring much about the separation on the second component which corresponds to the weak channel.

It is also observed that both the X-, Y-Precoders appear to adapt well to the changes in the channel. However, the Y-Precoders perform better than X-Precoders for $\beta \geq 3$, and hence for channels which are badly conditioned Y-Precoders would have a better error performance compared to X-Precoder. We shall see later that, indeed for the Rayleigh fading channel, Y-Precoders perform better than X-Precoder.

Therefore, we can conclude that codes which are fixed and do not change with each channel realization would have a poor error performance for large values of $\beta$ since they would waste power along the second component without any effect on the effective Euclidean distance. In fading channels, $\beta$ can be very large at times. Therefore, , a good code is one which adapts to $\beta$.

We also observe that Y-Codes and Y-Precoders have the best error performance when channel condition is bad. This justifies the fact that codes for badly conditioned channels should be designed to have more separation in the minimum distance along the component corresponding to the stronger channel.

### B. Diversity order comparison

We next discuss the diversity order achieved by the various precoding schemes with Rayleigh fading. Let the number of subchannels used for transmission be $n_s$ ($n_s \leq n_r$). The diversity order achieved by the linear precoders (EE and ARITH-MBER) and THP is $(n_r - n_s + 1)(n_t - n_s + 1)$ and $(n_t - n_s + 1)$ respectively, whereas the diversity order achieved by E-dmin and X-, Y-Codes is $(n_r - \frac{n_s}{2} + 1)(n_t - \frac{n_s}{2} + 1)$. The CI scheme achieves infinite diversity, but it suffers from power enhancement at the transmitter. Among all the other schemes (except CI), we observe that E-dmin and X-, Y-Codes have the best diversity order. The subsequent simulation results assume a Rayleigh fading channel.

### C. Comparison of BER performance with full-rate transmission

In Fig. 6, we plot the bit error rate (BER) of all precoders for $n_r = n_t = n_s = 2, 4$ and a target spectral efficiency of $2n_s$ bps/Hz. The proposed X, Y-Precoders and E-dmin have the best error performance. The increased diversity order achieved by the pairing scheme is obvious from the higher slope of the error rate for the X, Y-Precoders compared to a slope of order 1 for the linear precoder ARITH-MBER and THP. The performance of CI is inferior due to enhanced transmit power requirement arising from the bad conditioning of the channel. It is observed that the proposed Y-Precoders perform the best for $n_r = n_t = 2$, with E-dmin only $0.5$ dB away at BER of $10^{-3}$. For $n_r = n_t = 4$, E-dmin performs better than Y-Precoders by $0.4$ dB at BER of $10^{-3}$. However, E-dmin has this performance gain at a higher encoding and decoding complexity compared to the Y-Precoder.

### D. Comparison of BER performance for $n_r = n_t = 2, 4$

In Fig. 7, we plot the BER for $n_r = n_t = 2$, and a target spectral efficiency of $4, 8$ bps/Hz. It is observed that the best performance is achieved by the proposed Y-Precoder. For a target spectral efficiency of $4$ bps/Hz, ARITH-MBER also has a similar performance. However, for a spectral efficiency of $8$ bps/Hz, the performance of ARITH-MBER is worse than that of Y-Precoders by about $2.8$ dB at a BER of $10^{-3}$. This is because, to achieve higher diversity order, linear precoders do not use all the modes of transmission (i.e. $n_s < \min(n_r, n_t)$). Hence, to achieve the same target spectral efficiency, they have to use higher order QAM, which results in loss of power efficiency.

In Fig. 8, we plot the BER for $n_r = n_t = 4$, and a target spectral efficiency of $8, 16$ bps/Hz. For a target spectral efficiency of $8$ bps/Hz, E-dmin and ARITH-MBER have the best error performance. Y-Precoders perform only about $0.5$ dB away at a BER of $10^{-3}$. However, for a target spectral efficiency of $16$ bps/Hz Y-Precoders perform the best. ARITH-MBER ($n_s = 2$ with 256-QAM modulation on both channels) performs $2.6$ dB worse than Y-Precoders at a BER of $10^{-3}$. E-dmin performs the worst and is about $3.5$ dB away from Y-Precoders at BER of $10^{-3}$. E-dmin has poor performance since the precoder proposed in [12] has been optimized for 4-QAM, and therefore, it does not perform that well when the target spectral efficiency is higher than $2n_t$ bps/Hz.

### E. X-Codes vs. Y-Codes

In Fig. 9, we compare the BER performance of the proposed X and Y-Codes for a $n_r = n_t = 2$ system with spectral efficiency of $4, 8$ bps/Hz. It is observed that Y-Codes have a significant performance gain over X-Codes. For a target spectral efficiency of both $4$ and $8$ bps/Hz, Y-Codes perform better than X-Codes by about $1.5$ dB at a BER of $10^{-3}$. This is primarily due to the novel constellation structure of the proposed Y-Codes (as compared to the simple rotation encoder for X-Codes),which ensures that the minimal distance between constellation points does not become too small when channel is poorly conditioned. In Fig. 10, we compare the BER performance of the proposed X and Y-Codes for a $n_r = n_t = 4$ system with spectral efficiency of $8, 16$ bps/Hz. Y-Codes again perform better than X-Codes by about $0.7$ dB for a spectral efficiency of $8$ bps/Hz, and by about $1.5$ dB for a spectral efficiency of $16$ bps/Hz.

### F. X-, Y-Codes vs. X-, Y-Precoders

In this section, we discuss the performance gain achieved by optimally choosing the encoder matrices $\mathbf{A}_k$ for each channel realization, as compared to having them fixed *a priori*.

In Fig. 9, we compare the performance of the X-, Y-Precoders with that of X-, Y-Codes for $n_r = n_t = 2$ with a target spectral efficiency of $4, 8$ bps/Hz. For a target spectral efficiency of $4$ bps/Hz, X-, Y-Precoders perform only marginally better than X-, Y-Codes (by only about $0.2$ dB at BER of $10^{-3}$). However, for a target spectral efficiency of 8 bps/Hz, the X-Precoder performs better than X-Codes by about $1.0$ dB, whereas Y-Precoders perform better than Y-Codes by about $0.2$ dB at a BER of $10^{-3}$. Therefore, changing the encoder matrices with channel realization is beneficial for X-Codes. However, it is observed that Y-Precoders do not have as much gain in performance compared to Y-Codes.

For $n_r = n_t = 4$, it is observed from Fig. 10 that for a target spectral efficiency of 8 bps/Hz, X-, Y-Precoders have almost similar performance as X-, Y-Codes. However, for a target spectral efficiency of 16 bps/Hz, X-Precoders perform better than X-Codes by about $0.7$ dB, whereas Y-Precoders perform better than Y-Codes by about $0.3$ dB at a BER of $10^{-3}$.

The performance gain of X-Precoders over X-Codes is much more significant as compared to that of the Y-Precoders over Y-Codes. Also, for X-Precoders, this performance gain is significant only with higher order QAM. This is due to the fact that the error performance is much more sensitive to the rotation angle for higher order QAM (see Fig. 2), and therefore adjusting the rotation angle with respect to the varying channel is expected to result in performance improvement.

On the other hand Y-Precoders are only marginally better than Y-Codes irrespective of the spectral efficiency. This is attributed to the fact that for the Y-Precoders we optimize the upper bound to the probability of error rather than the exact error probability. We do this, because of the analytical intractability of the exact error probability expression. This leads to a suboptimal choice of the encoder matrices, and therefore a suboptimal error performance.

This is obvious from Fig. 11, where we plot the exact optimal word error probability in comparison with the error probability of the proposed suboptimal Y-Precoder. The exact optimal word error probability (i.e, error probability with the optimal choice of encoder matrices) is computed through Monte Carlo techniques using (53) and the integrals in (54). The exact optimal error probability is better than the proposed suboptimal Y-Precoder by about $1.8$ dB for a $n_r = n_t = 2$ system, and is better by about $1.0$ dB for a $n_r = n_t = 4$ system at a word error probability of $10^{-1}$. This therefore suggests the existence of better Y-Precoders compared to what has been proposed in this paper.

## VII. COMPLEXITY

In this section, we discuss the computational complexity of X-, Y-Codes and compare it with other precoding schemes. The linear precoders (ARITH-MBER and EE), E-dmin and X-Codes need to compute the SVD decomposition of $\mathbf{H}$. The CI and THP schemes involve computing the pseudo-inverse and QR decomposition of $\mathbf{H}$ respectively. The complexity of computing SVD, QR as well as pseudo-inverse is $O(n_r^3)$. These operations are computation intensive. However, TDD is generally employed in a slowly fading channel, and therefore these computations can be performed once, and can be used until the channel changes. We, therefore, do not consider the complexity of these decompositions in the discussion below.

## A. Encoding Complexity

The encoding complexity of all the schemes is $O(n_r n_t)$, which is due to the transmit preprocessing filter. If the number of operations were to be computed, CI and X-, Y-Codes would have the lowest complexity. This is so because, linear precoders need to compute an extra pre-processing matrix (in addition to SVD). THP also has to do successive interference pre-cancelation (in addition to QR). On the other hand, E-dmin and X-, Y-Codes need to only compute SVD, which automatically gives the pre-processing and the post-processing matrices. Also, X-, Y-Codes have lower encoding complexity compared to E-dmin, because the encoding matrices $\mathbf{A}_k$ are real, as opposed to being complex for E-dmin. CI has an even lower complexity since there is no spatial coding.

## B. Decoding Complexity

The decoding complexity of all the schemes have a square dependence on $n_r$. This is due to the post-processing matrix filter at the receiver. The linear precoders, CI and THP employ post processing at the receiver, which enables independent ML decoding for each subchannel. With QAM modulation symbols this is only a rounding operation for each subchannel. E-dmin and X-Codes on the other hand use sphere decoding to jointly decode pairs of subchannels. ML decoding for X-Codes is accomplished by using $n_r$ 2-dimensional real sphere decoders. However, E-dmin requires $\frac{n_r}{2}$ 4-dimensional real sphere decoders. The average complexity of sphere decoding is cubic in the number of dimensions (and is invariant w.r.t modulation alphabet size $M$) [13], and therefore X-Codes have a much lower decoding complexity when compared to E-dmin. The ML decoding complexity of Y-Codes is independent of $M$, and is equal to the ML decoding complexity of a scalar channel. Therefore, the linear precoders, CI, THP and Y-Codes have the lowest ML decoding complexity among the considered precoding schemes.

## VIII. Conclusion

We proposed X-, Y-Codes/Precoders which can achieve full-rate and high diversity at low complexity by pairing the subchannels prior to SVD precoding. It is observed that indeed pairing of channels can significantly improve the overall diversity. Among all possible pairings, pairing the $k$-th channel with the $(n_r - k + 1)$-th subchannel was found to be optimal in terms of achieving the best diversity order. One way of pairing the subchannels is by using rotation based encoding as for X-Codes/Precoders. The proposed X-Codes/Precoders have good performance for well conditioned channels. For ill-conditioned channels, we then proposed Y-Codes/Precoders. It is shown by simulation and analysis that Y-Codes/Precoders achieve the best error performance at very low complexity, when compared to other precoders in the literature. In practice, in order to improve the overall performance, it is possible to adaptively switch between X- and Y-Codes/Precoders depending on the channel conditions.

## Appendix A

### Proof of Theorem 1

Towards proving Theorem 1, we shall find the following Lemma useful (See proposition 1 in [17] for the same lemma and its proof).

*Lemma 1:* Given a real scalar channel modeled by $y = \sqrt{\alpha}x + n$, where $x = \pm\sqrt{E_s}$, $n \sim \mathcal{N}(0, \sigma^2)$. Let $F(\alpha) = C\alpha^k + o(\alpha^k)$, for $\alpha \rightarrow 0^+$ be the cdf (cumulative density function) of $\alpha$, where $C$ is a constant and $k$ is a positive integer[2]. Let $\gamma = E_s/\sigma^2$ be the SNR. Then the probability of error is given by $P(\gamma) = \mathbb{E}_\alpha[Q(\sqrt{\alpha\gamma})]$. The lemma states that the asymptotic error probability for $\gamma \rightarrow \infty$ is given by

$$P = \frac{C((2k-1)\cdot(2k-3)\;\cdots\;5\cdot3\cdot1)}{2}\gamma^{-k} + o(\gamma^{-k})$$

∎

The proof of Theorem 1 is as follows. Let $\{\Re(\mathbf{u}_k) \rightarrow \Re(\mathbf{v}_k)\}$ denote the pairwise error event that, given $\mathbf{u}_k$ was transmitted on the $k$-th pair, the real part of the ML detector for the $k$-th pair decodes in favor of some other vector $\Re(\mathbf{v}_k)$. Further, let us denote the probability of this event by $P_k^{'}(\Re(\mathbf{u}_k) \rightarrow \Re(\mathbf{v}_k))$ (pairwise error probability (PEP)). Using the union bounding technique, $P_k^{'}(\Re(\mathbf{u}_k))$ is then upper bounded by the sum of all the possible pairwise error probabilities. From (21) it is clear that this upper bound on $P_k^{'}(\Re(\mathbf{u}_k))$ induces an upper bound on $P_k^{'}$ which is given by

$$P_k^{'} \leq \frac{1}{|\mathcal{S}_k|}\sum_{\Re(\mathbf{u}_k)}\sum_{\Re(\mathbf{v}_k)\neq\Re(\mathbf{u}_k)}P_k^{'}(\Re(\mathbf{u}_k) \rightarrow \Re(\mathbf{v}_k)) \tag{60}$$

Due to Gaussian noise, this can be further written as

$$P_k^{'} \leq \frac{1}{|\mathcal{S}_k|}\sum_{\Re(\mathbf{u}_k)}\sum_{\Re(\mathbf{v}_k)\neq\Re(\mathbf{u}_k)}\mathbb{E}\left[Q\left(\sqrt{\frac{d_k^2(\Re(\mathbf{u}_k),\Re(\mathbf{v}_k),\mathbf{A}_k)}{2N_0}}\right)\right] \tag{61}$$

where

$$d_k^2(\Re(\mathbf{u}_k),\Re(\mathbf{v}_k),\mathbf{A}_k) \;\triangleq\; \|\mathbf{M}_k(\Re(\mathbf{u}_k) - \Re(\mathbf{v}_k))\|^2$$

$$Q(x) \;\triangleq\; \int_x^\infty \frac{1}{\sqrt{2\pi}}e^{-t^2/2}dt$$

The expectation in (61) is over the joint distribution of the channel gain $(\lambda_{i_k}, \lambda_{j_k})$. The joint pdf (probability density function) of the ordered eigenvalues of $\mathbf{H}^\dagger\mathbf{H}$ is given by the well known Wishart distribution [14]. However, in (61) evaluating the expectation over $(\lambda_{i_k}, \lambda_{j_k})$ is still a difficult problem except for trivial cases (like $n_r = n_t = 2$). We therefore try to bound $d_k^2(\Re(\mathbf{u}_k),\Re(\mathbf{v}_k),\mathbf{A}_k)$ such that the bound depends only on $\lambda_{i_k}$. Since $\lambda_{i_k} \geq \lambda_{j_k} \geq 0$, using the definition of $\mathbf{M}$ and $\mathbf{M}_k$, we have

$$d_k^2(\Re(\mathbf{u}_k),\Re(\mathbf{v}_k),\mathbf{A}_k) \geq \lambda_{i_k}^2 \tilde{d}_k^2(\Re(\mathbf{u}_k),\Re(\mathbf{v}_k),\mathbf{A}_k) \tag{62}$$

where

$$\tilde{d}_k^2(\Re(\mathbf{u}_k),\Re(\mathbf{v}_k),\mathbf{A}_k) \;\triangleq\; e_{k,1}^2$$

$$\mathbf{e_k} \;\triangleq\; \mathbf{A}_k(\Re(\mathbf{u}_k) - \Re(\mathbf{v}_k))$$

---

[2] Any function $f(x)$ in a single variable $x$ is said to be $o(g(x))$ i.e. $f(x) = o(g(x))$ if $\frac{f(x)}{g(x)} \rightarrow 0$ as $x \rightarrow 0$

and we let $e_{k,1}$ denote the first component of the 2-dimensional vector $\mathbf{e_k}$. We further define the *generalized minimum distance* as follows :

$$g_k(\mathbf{A}_k) = \min_{\Re(\mathbf{u}_k) \neq \Re(\mathbf{v}_k)} \tilde{d}_k^2(\Re(\mathbf{u}_k), \Re(\mathbf{v}_k), \mathbf{A}_k) \tag{63}$$

Since $Q(\cdot)$ is a monotonically decreasing function with increasing argument, we can further upper bound (61) using (63) as follows :

$$P_k^{'} \leq (|\mathcal{S}_k| - 1)\mathbb{E}\left[ Q\left( \sqrt{\frac{\lambda_{i_k}^2 g_k(\mathbf{A}_k)}{2N_0}} \right) \right] \tag{64}$$

For a Rayleigh faded channel, the marginal pdf of the $s$-th eigenvalue $\lambda_s^2$ (for $\lambda_s^2 \to 0$) is given by [16]

$$f(\lambda_s^2) = C(s)(\lambda_s^2)^{N_t(s)N_r(s)-1} + o\left( (\lambda_s^2)^{N_t(s)N_r(s)-1} \right) \tag{65}$$

where $N_t(s) \triangleq (n_t - s + 1)$, $N_r(s) \triangleq (n_r - s + 1)$ and $C(s)$ is a constant given in [16]. Using the pdf in (65), the cdf $F_s(u) = P(\lambda_s^2 \leq u)$ (for $u \to 0^+$) is given by

$$F_s(u) = D(s)u^{N_t(s)N_r(s)} + o\left( u^{N_t(s)N_r(s)} \right) \tag{66}$$

where $D(s) \triangleq \frac{C(s)}{N_t(s)N_r(s)}$. Using Lemma 1 and (66), the bound in (64) can be further written as

$$P_k^{'} \leq (|\mathcal{S}_k| - 1)c_k\left( \frac{\gamma g_k(\mathbf{A}_k)}{2P_T} \right)^{-\delta_k} + o(\gamma^{-\delta_k}) \tag{67}$$

where $\delta_k \triangleq (n_t - i_k + 1)(n_r - i_k + 1)$ and $c_k \triangleq \frac{C(i_k)((2\delta_k - 1) \cdot \ \cdots \ 5 \cdot 3 \cdot 1)}{2\delta_k}$. ∎

# Appendix B

## Proof of Theorem 2

Let $d(\theta_k, \lambda_{i_k}, \lambda_{j_k}) \triangleq \min_{(p,q) \in \mathbb{S}_2} d_k^2(p, q, \theta_k)$, where $d_k^2(p, q, \theta_k)$ is defined in (40). The objective is to find the optimal $\theta_k$ which maximizes $d(\theta_k, \lambda_{i_k}, \lambda_{j_k})$. The set $\mathbb{S}_2$ contains exactly 8 elements. Also due to sign symmetry of this set ( i.e. if $(p, q) \in \mathbb{S}_2$ then so do $(p, -q)$, $(-p, q)$ and $(-p, -q)$), there are actually only 4 distances to be computed. For a given angle $\theta_k$ these distances are enumerated as follows :

$$
\begin{aligned}
d_1 &= \lambda_{i_k}^2 \cos^2(\theta_k) + \lambda_{j_k}^2 \sin^2(\theta_k) \\
d_2 &= \lambda_{i_k}^2 \sin^2(\theta_k) + \lambda_{j_k}^2 \cos^2(\theta_k) \\
d_3 &= \lambda_{i_k}^2 (\cos(\theta_k) + \sin(\theta_k))^2 + \lambda_{j_k}^2 (\cos(\theta_k) - \sin(\theta_k))^2 \\
d_4 &= \lambda_{i_k}^2 (\cos(\theta_k) - \sin(\theta_k))^2 + \lambda_{j_k}^2 (\cos(\theta_k) + \sin(\theta_k))^2
\end{aligned}
$$

Therefore, $d(\theta_k, \lambda_{i_k}, \lambda_{j_k})$ can be expressed in terms of these distances as

$$
\begin{aligned}
d(\theta_k, \lambda_{i_k}, \lambda_{j_k}) &= \min(d_1, d_2, d_3, d_4) \\
&= \min(\min(d_1, d_2), \min(d_3, d_4))
\end{aligned}
$$

Just as in Appendix A, it can be shown that for the maximization in (39), it suffices to only consider the range of $\theta_k$ to be $[0, \pi/4]$. Also it is given that $\lambda_{i_k} \geq \lambda_{j_k}$ due to the order of the singular values in the SVD decomposition and the way in which pairing is done for X-Codes. Using these two facts, it can be concluded that

$$d_2 = \min(d_1, d_2)$$

$$d_4 = \min(d_3, d_4)$$

and therefore

$$d(\theta_k, \lambda_{i_k}, \lambda_{j_k}) = \min(d_2, d_4)$$

Let

$$\beta_k \triangleq \frac{\lambda_{i_k}}{\lambda_{j_k}}$$

Then, $d_4$ is the minimum if the following condition is satisfied.

$$\frac{1}{\beta_k^2} \leq h(\theta_k) \tag{68}$$

where

$$h(\theta_k) \triangleq 1 - \frac{1}{\sin(2\theta_k) + \sin^2(\theta_k)}$$

It can be seen that over the interval $(0 \ \frac{\pi}{4}]$, $h(\theta_k)$ is a continuous and monotonically increasing function. This is because the first derivative w.r.t $\theta_k$ exists and is always positive. The maximum value of $h(\theta_k)$ over this interval is $\frac{1}{3}$. Therefore, we now consider two situations depending upon whether $\beta_k$ is greater than or less than $\sqrt{3}$.

If $\beta_k \leq \sqrt{3}$, then $\frac{1}{\beta_k^2} \geq \frac{1}{3}$. Since $h(\theta_k)$ is always less than $\frac{1}{3}$, we can conclude that the condition in (68) is never satisfied and therefore $d_2$ is the minima. Further since $\lambda_{i_k} \geq \lambda_{j_k}$, $d_2$ is a monotonically increasing function of $\theta_k$ and therefore the solution to the maximization problem in (39) is $\frac{\pi}{4}$.

If $\beta_k \geq \sqrt{3}$, then $\frac{1}{\beta_k^2} \leq \frac{1}{3}$. Since $h(\theta_k)$ is a monotonically increasing function we observe that $d_4$ is the minima when $\theta_k \geq \theta_k^*$ or else $d_2$ is the minima. Here $\theta_k^*$ is such that $\theta_k^* \in [0 \ \frac{\pi}{4}]$ and $h(\theta_k^*) = \frac{1}{\beta_k^2}$. Further, it is observed that $d_2$ is a monotonically increasing function of $\theta_k$ whereas $d_4$ is monotonically decreasing. Also, $d_2 = d_4$ when $\theta_k = \theta_k^*$. Therefore, it can be concluded that $\min(d_2, d_4)$ is maximized when $\theta_k = \theta_k^*$. Hence for $\beta_k \geq \sqrt{3}$ the solution to the maximization problem in (39) is $\theta_k^*$. We now solve for $\theta_k^*$. Using the definition of $h(\theta_k)$, we have

$$\frac{1}{\beta_k^2} = 1 - \frac{1}{\sin(2\theta_k^*) + \sin^2(\theta_k^*)}$$

or equivalently,

$$\tan^2(\theta_k^*) - 2(\beta_k^2 - 1)\tan(\theta_k^*) + \beta_k^2 = 0$$

The last equation is quadratic in $\tan(\theta_k^*)$, and the solution which results in $\theta_k^* \in [0 \ \frac{\pi}{4}]$ is given by

$$\theta_k^* = \tan^{-1}\left[(\beta_k^2 - 1) - \sqrt{((\beta_k^2 - 1)^2 - \beta_k^2)}\right]$$

Combining, the optimal angles obtained for $\beta_k \leq \sqrt{3}$ and $\beta_k > \sqrt{3}$, we get the solution to the maximization problem as stated in (41). ∎

## APPENDIX C

## PROOF OF THEOREM 3

We first get an expression for $d_{k,\min}^2(a_k, b_k)$ as defined in (56). For any code vector at index $v$ which is even, the nearest distance to any other code vector with even index is $4\lambda_{i_k}^2 a_k^2$. The nearest distance to any code vector at odd index is $\lambda_{i_k}^2 a_k^2 + 4\lambda_{j_k}^2 b_k^2$. The same holds true if $v$ is odd. Hence $d_{k,\min}^2(a_k, b_k)$ is given by

$$d_{k,\min}^2(a_k, b_k) = \min\left(4\lambda_{i_k}^2 a_k^2, \lambda_{i_k}^2 a_k^2 + 4\lambda_{j_k}^2 b_k^2\right) \tag{69}$$

Therefore, our objective is to solve the following constrained min-max optimization problem.

$$(a_k^*, b_k^*) = \arg \max_{\{(a_k, b_k) \in \mathbb{R}^{2+} | b_k^2 + a_k^2 \frac{M^2-1}{12} = \frac{P_T}{n_r}\}} \min\left(4\lambda_{i_k}^2 a_k^2, \lambda_{i_k}^2 a_k^2 + 4\lambda_{j_k}^2 b_k^2\right) \tag{70}$$

In (70), we let $T_1 \triangleq 4\lambda_{i_k}^2 a_k^2$, which is geometrically a straight line w.r.t. $a_k^2$ passing through the origin and attaining a maximum value of

$$\frac{48 P_T \lambda_{i_k}^2}{n_r(M^2-1)}$$

at

$$a_k^2 = \frac{12 P_T}{n_r(M^2-1)}$$

This is due to the transmit power constraint

$$a_k^2 \leq \frac{12 P_T}{n_r(M^2-1)}$$

In (70), we let $T_2 \triangleq (\lambda_{i_k}^2 a_k^2 + 4\lambda_{j_k}^2 b_k^2)$. Since

$$b_k^2 + a_k^2 \frac{(M^2-1)}{12} = \frac{p_T}{n_r}$$

we can express $T_2$ as

$$T_2 = \lambda_{j_k}^2\left(4\frac{P_T}{n_r} + a_k^2\left(\beta_k^2 - \frac{M^2-1}{3}\right)\right) \tag{71}$$

From (71), we observe that, if $\beta_k^2 \geq (M^2-1)/3$, then $T_2$ is a straight line with positive slope, with a value of $4\lambda_{j_k}^2 P_T/n_r$ at $a_k = 0$ and attaining a maximum value of

$$\frac{12 P_T \lambda_{i_k}^2}{n_r(M^2-1)}$$

at

$$a_k^2 = \frac{12 P_T}{n_r(M^2-1)}$$

This maximum value is less than the maximum attained by $T_1$. Since both $T_1$ and $T_2$ have positive slope the minimum among $T_1$ and $T_2$ is maximized at $a_k^{*^2} = 12P_T/(n_r(M^2-1))$, which implies that $b_k^* = 0$. The value of $d_{k,\min}^2(a_k, b_k)$ at $a_k = a_k^*$ is the maximum value attained by $T_2$. Therefore, when the channel condition exceeds a certain threshold, it is optimal to allocate all power to the stronger channel only.

On the other hand, if $\beta_k^2 < \frac{M^2-1}{3}$, then $T_2$ is a straight line with negative slope, whereas $T_1$ has positive slope, and therefore the minimum between them is maximized when they are both equal. Therefore, the optimal $(a_k^*, b_k^*)$ must satisfy

$$4\lambda_{i_k}^2 a_k^{*2} = \lambda_{i_k}^2 a_k^{*2} + 4\lambda_{j_k}^2 b_k^{*2} \tag{72}$$

Using the fact that $b_k^{*2} + a_k^{*2} \frac{M^2-1}{12} = \frac{P_T}{n_r}$, the optimal $(a_k^*, b_k^*)$ is given by

$$(a_k^*, b_k^*) = \left( \frac{\sqrt{\frac{4P_T}{3n_r}}}{\sqrt{\beta_k^2 + \frac{M^2-1}{9}}}, \sqrt{\frac{P_T}{n_r}} \frac{\beta_k}{\sqrt{\beta_k^2 + \frac{M^2-1}{9}}} \right) \tag{73}$$

Using (73), the optimal value of $d_{k,\min}^2(a_k, b_k)$ for $\beta_k^2 < \frac{M^2-1}{3}$ is given by

$$d_{k,\min}^2(a_k^*, b_k^*) = \frac{16P_T \lambda_{i_k}^2}{n_r \left( 3\beta_k^2 + \frac{(M^2-1)}{3} \right)}$$

∎

## References

[1] G. Raleigh and J. Cioffi, "Spatio-Temporal Coding for Wireless Communication", *IEEE Trans. Commun.*, pp. 357–366, March 1998.

[2] R. Knopp and G. Caire, "Power Control Schemes for TDD Systems with Multiple Transmit and Receive Antennas", *Proc. of IEEE Global Telecommunications Conference (Globecom)*, pp. 2326–2330, Rio de Janeiro, Nov. 1999.

[3] T.M. Cover and Joy A. Thomas, *Elements of information theory*, John Wiely and Sons, 2nd Ed., July 2006.

[4] P.W. Baier, M. Meurer, T. Weber, and H. Troeger, "Joint Transmission (JT), an alternative rationale for the downlink of Time Division CDMA using multi-element transmit antennas", *Proc. of IEEE Int. Symp. on Spread Spectrum Techniques and Applications (ISSSTA)*, pp. 1–5, Parsippany, NJ, Sep. 2000.

[5] H. Harashima, and H. Miyakawa, "Matched transmission technique for channels with inter-symbol interference", *IEEE Trans. on Communications*, vol. 20, pp. 774–780, 1972.

[6] M. Tomlinson, "New automatic equaliser employing modulo arithmetic", *Electronics Letters*, 7, pp. 138–139, 1971.

[7] R.F.H. Fischer, C. Windpassinger, A. Lampe, and J.B. Huber, "Space-Time Transmission using Tomlinson-Harashima Precoding", *Proc. Int. Zurich Seminar on Broadband Communications (IZS'02)*, Zurich, Switzerland, Feb. 2002.

[8] D. Perez Palomar, J.M. Cioffi, and M.A. Lagunas, "Joint Tx-Rx Beamforming Design for Multicarrier MIMO Channels: A unified Framework for Convex Optimization", *IEEE. Trans. on Signal Processing*, pp. 2381–2401, vol. 51, no. 9, Sept. 2003.

[9] A. Scaglione, P. Stoica, S. Barbarossa, and G.B. Giannakis, Hemanth Sampath, "Optimal Designs for Space-Time Linear Precoders and Decoders", *IEEE Trans. on Signal Processing*, pp. 1051–1064, vol. 50, no. 5, May. 2002.

[10] C. Windpassinger and R.F.H. Fischer, "Low-Complexity Near-Maximum-Likelihood Detection and Precoding for MIMO Systems using Lattice Reduction", *Proc. of Information Theory Workshop (ITW '03)*, Paris, France, March 31 - April 4, 2003.

[11] C.B. Peel, B.M. Hochwald, and A.L. Swindlehurst, "A vector-perturbation technique for near-capacity multiantenna multiuser communication-part I: channel inversion and regularization", *IEEE Trans. on Communications*, vol. 53, no. 1, Jan. 2005, pp. 195–202.

[12] B. Vrigneau, J. Letessier, P. Rostaing, L. Collin, and G. Burel, "Extension of the MIMO Precoder Based on the Minimum Euclidean Distance: A Cross-Form Matrix", *IEEE J. of Selected Topics in Signal Processing*, pp. 135–146, vol. 2, no. 2, April. 2008.

[13] B. Hassibi and H. Vikalo, "On the Sphere-Decoding Algorithm I. Expected Complexity", *IEEE Trans. on Information Theory*, vol. 53, no. 8, Aug. 2005.

[14] A.M. Tulino and S. Verdú, "Random Matrix Theory and Wireless Communications", *Foundations and Trends in Communications and Information Theory*, vol. 1, no. 1.

[15] E. Viterbo and J. Boutros, "A Universal Lattice Code Decoder for Fading Channels", *IEEE Transactions on Information Theory*, vol. 45, n. 5, pp. 1639-1642, July 1999.

[16] L.G. Ordonez, D.P. Palomar, A.P. Zamora, and J.R. Fonollosa, "High-SNR analytical performance of Spatial Multiplexing MIMO Systems With CSI", *IEEE Trans. on Signal Processing*, pp. 5447–5463, vol. 55, no. 11, Nov. 2007.

[17] Z. Wang and G.B. Giannakis, "A Simple and General Parametrization Quantifying Performance in Fading Channels", *IEEE Trans. on Communications*, pp. 1389–1398, vol. 51, no. 8, Aug. 2003.

**Figure Captions:**

1) Union bound for word error probability. $n_r = n_t = 2$ and $M = 2$ (4-QAM) modulation.

2) Sensitivity of word error probability w.r.t $\theta_1$. $n_r = n_t = 2$ and $M = 2,4$ (4,16-QAM) modulation.

3) One quadrant of the set $\mathbb{S}_M$ for $M = 2,4$ (4,16-QAM modulation). The critical angles where performance degrades severely are shown to coincide with $\tan^{-1}(-p/q)$.

4) Received signal space for the real component of the $k$-th pair. $M = 8$ and therefore we have 5 regions with vertical dashed lines demarcating the boundary between the regions. The scaled codebook vectors are represented by small filled circles along with their corresponding codebook index number. Dotted lines demarcate the boundary between the ML decision regions.

5) Effect of the channel condition number on error performance of various precoders for a $2 \times 2$ system with target spectral efficiency of 8bps/Hz.

6) BER comparison between various precoders for $n_r = n_t = n_s = 2, 4$ and $M = 2$ (4-QAM) modulation. Target spectral efficiency is equal to $2n_s$ bps/Hz.

7) BER comparison between various precoders for $n_r = n_t = 2$ and target spectral efficiency of 4, 8 bps/Hz.

8) BER comparison between various precoders for $n_r = n_t = 4$ and target spectral efficiency of 8, 16 bps/Hz.

9) BER comparison between the proposed X-Codes and Y-Codes for $n_r = n_t = 2$ with spectral efficiency = 4, 8 bps/Hz.

10) BER comparison between the proposed X-Codes and Y-Codes for $n_r = n_t = 4$ with spectral efficiency = 8, 16 bps/Hz.

11) Word error probability comparison between the proposed suboptimal Y-Precoders and exact optimal Y-Precoders for $n_r = n_t = 2, 4$ with spectral efficiency of $4n_t$ bps/Hz.

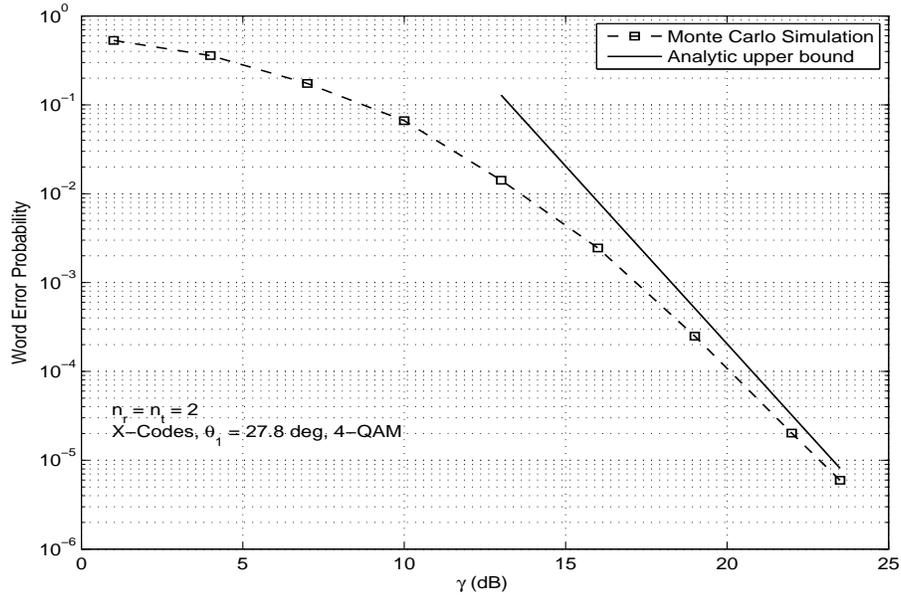

Fig. 1. Union bound for word error probability. $n_r = n_t = 2$ and $M = 2$ (4-QAM) modulation.

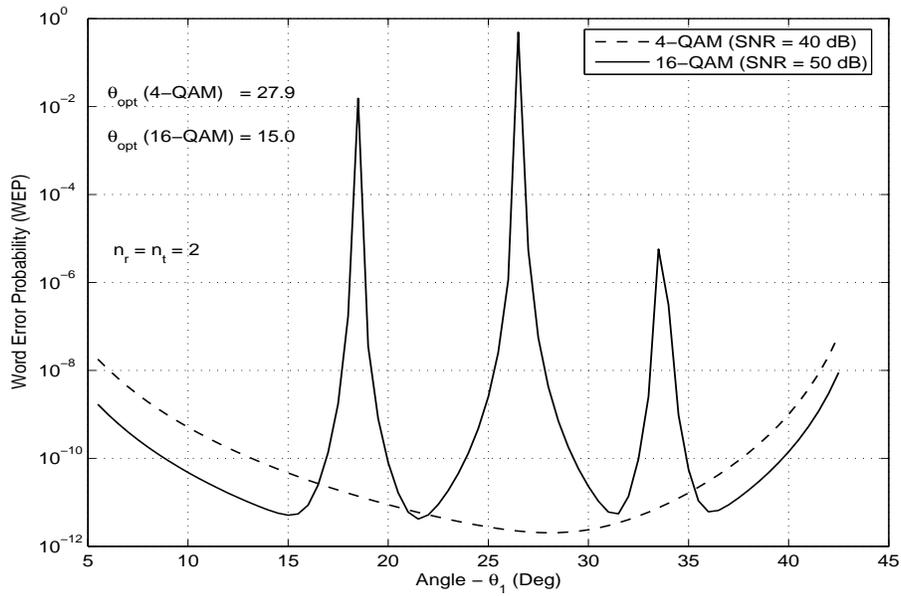

Fig. 2. Sensitivity of word error probability w.r.t $\theta_1$. $n_r = n_t = 2$ and $M = 2,4$ (4,16-QAM) modulation.

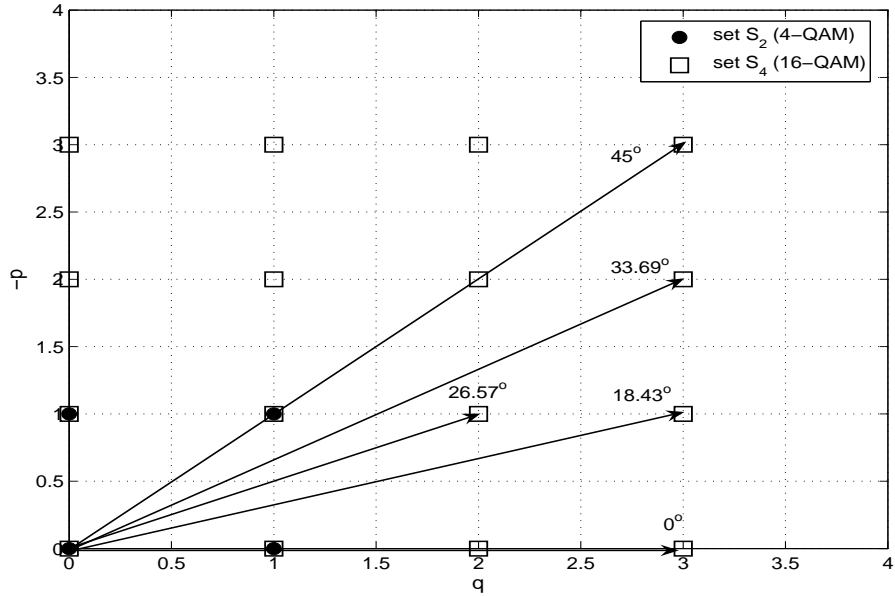

Fig. 3. One quadrant of the set $\mathbb{S}_M$ for $M = 2,4$ (4,16-QAM modulation). The critical angles where performance degrades severely are shown to coincide with $\tan^{-1}(-p/q)$.

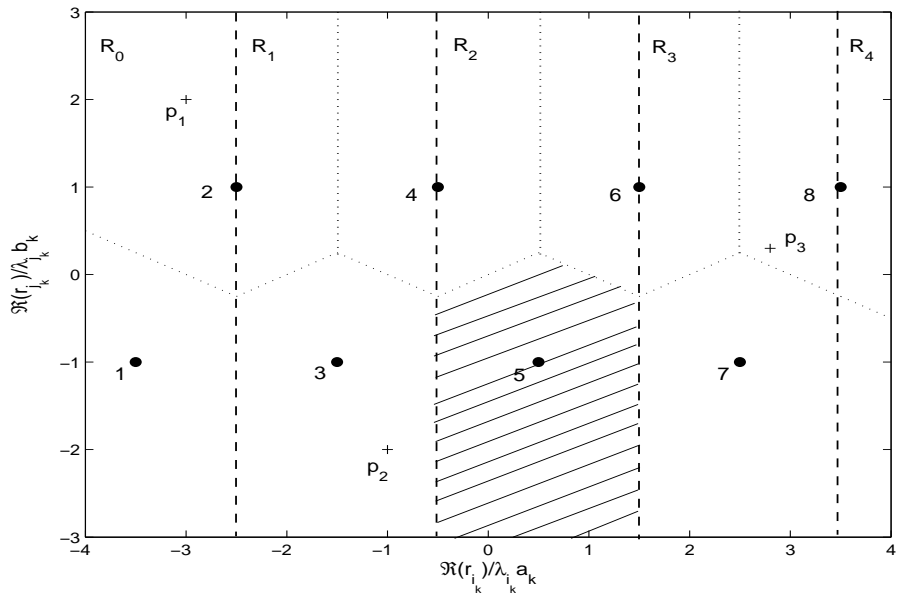

Fig. 4. Received signal space for the real component of the $k$-th pair. $M = 8$ and therefore we have 5 regions with vertical dashed lines demarcating the boundary between the regions. The scaled codebook vectors are represented by small filled circles along with their corresponding codebook index number. Dotted lines demarcate the boundary between the ML decision regions.

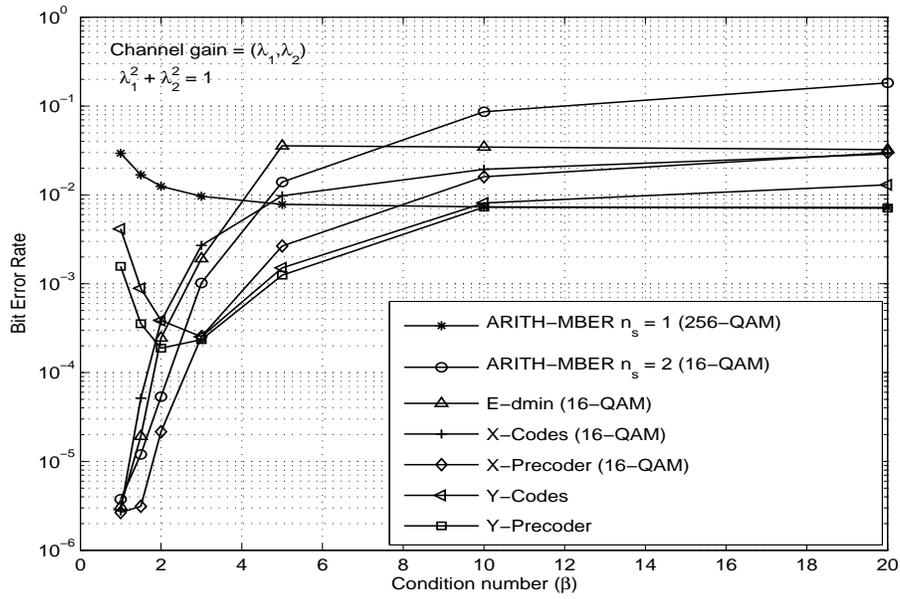

Fig. 5. Effect of the channel condition number on error performance of various precoders for a $2 \times 2$ system with target spectral efficiency equal to 8bps/Hz.

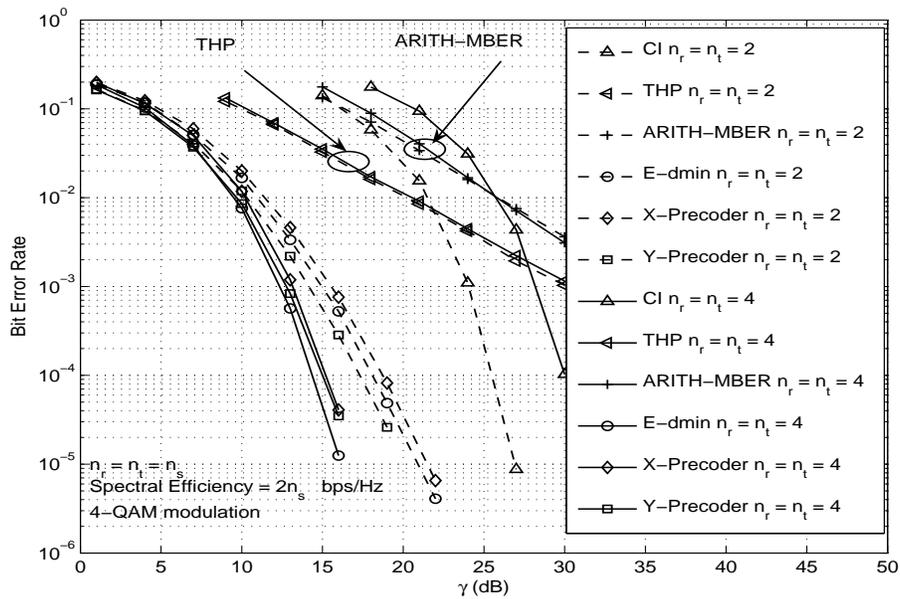

Fig. 6. BER comparison between various precoders for $n_r = n_t = n_s = 2, 4$ and $M = 2$ (4-QAM) modulation. Target spectral efficiency is equal to $2n_s$ bps/Hz.

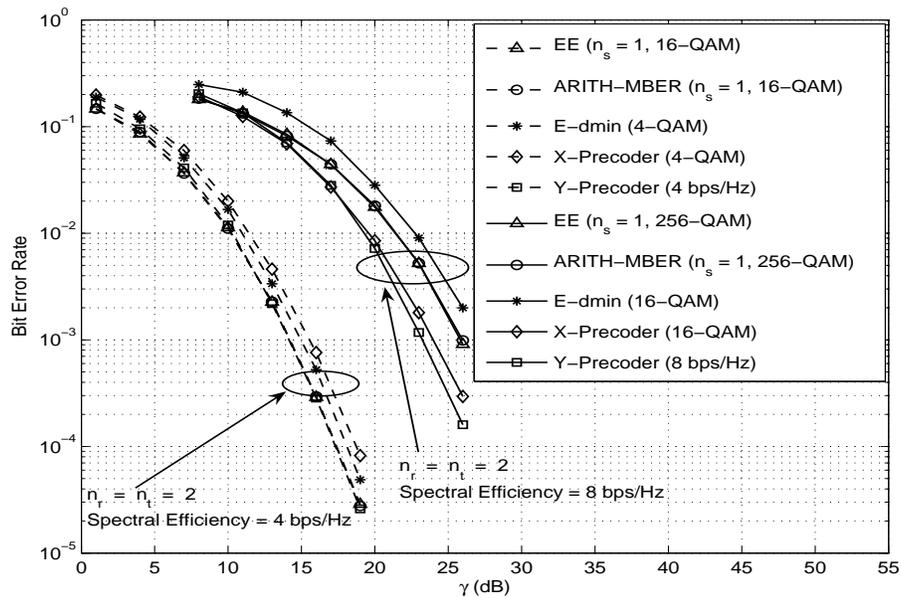

Fig. 7. BER comparison between various precoders for $n_r = n_t = 2$ and target spectral efficiency of 4, 8 bps/Hz.

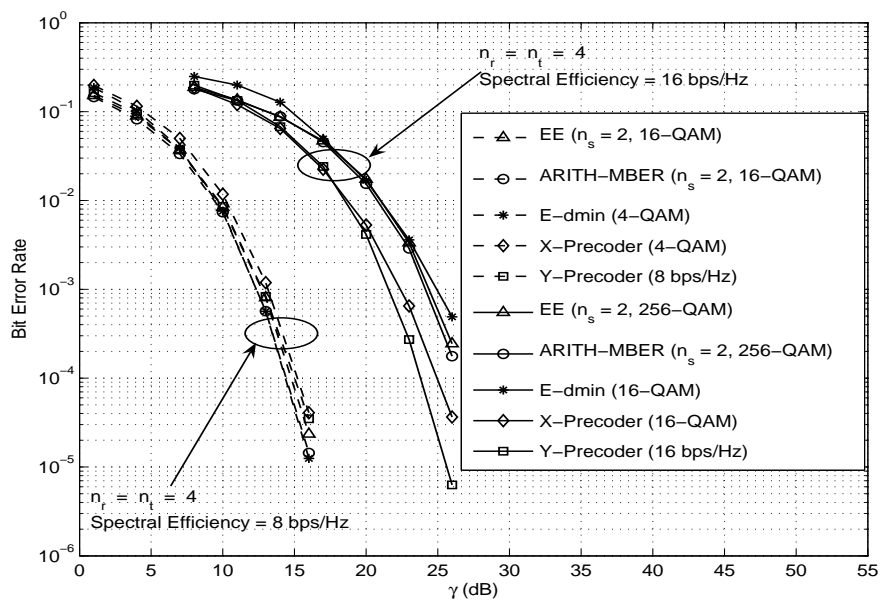

Fig. 8. BER comparison between various precoders for $n_r = n_t = 4$ and target spectral efficiency of 8, 16 bps/Hz.

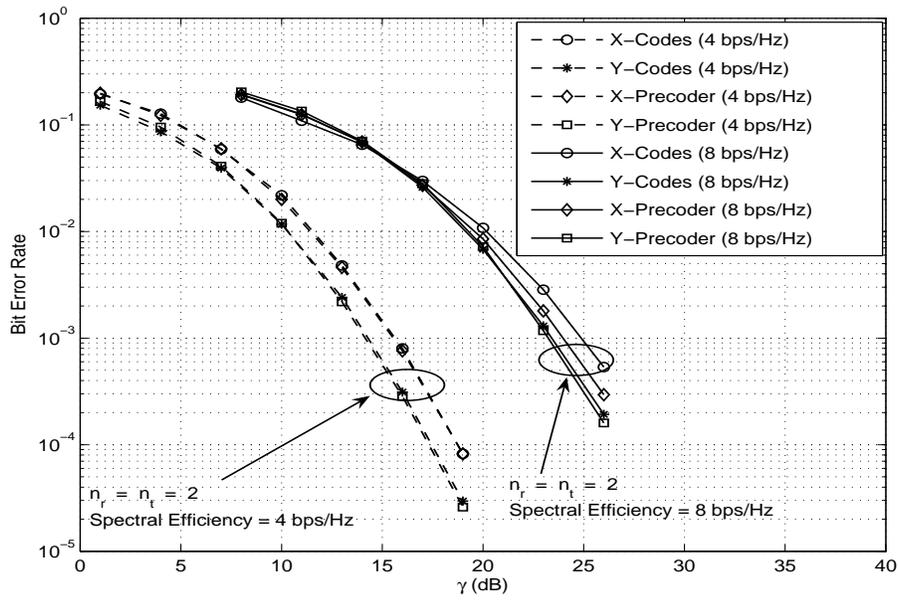

Fig. 9.   BER comparison between the proposed X-Codes and Y-Codes for $n_r = n_t = 2$ with spectral efficiency of 4, 8 bps/Hz.

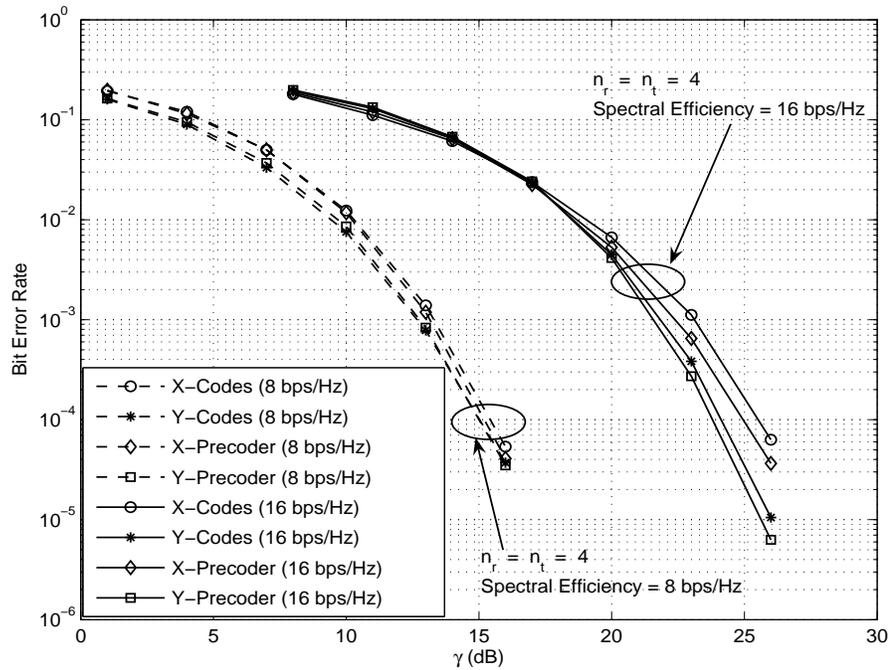

Fig. 10.   BER comparison between the proposed X-Codes and Y-Codes for $n_r = n_t = 4$ with spectral efficiency of 8, 16 bps/Hz.

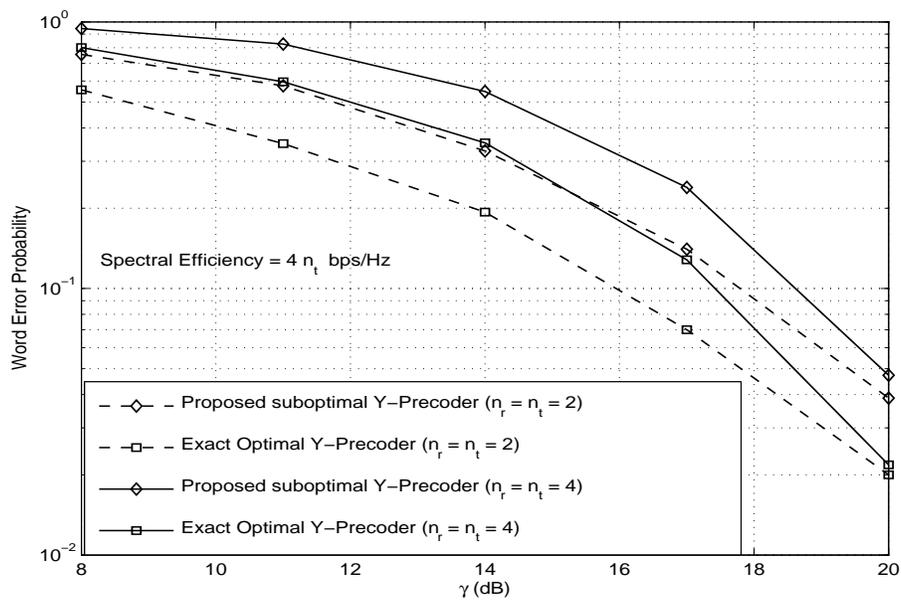

Fig. 11. Word error probability comparison between the proposed suboptimal Y-Precoders and exact optimal Y-Precoders for $n_r = n_t = 2, 4$ with spectral efficiency of $4n_t$ bps/Hz.